\documentclass[12pt]{article}
\usepackage{arxiv}
\usepackage{graphicx} 
\usepackage{tikz}
\usepackage{url}
\usepackage{booktabs}
\usepackage{subcaption}
\usepackage{adjustbox}
\usepackage{multirow,multicol}

\usepackage{amsmath,amsfonts}
\usepackage{amsthm}
\newtheorem{lemma}{Lemma}
\usepackage{pifont}
\usepackage{makecell}

\newcommand{\xmark}{\textcolor{red}{\ding{55}}}
\newcommand{\cmark}{\textcolor{green!60!black}{\ding{51}}}
\newcommand{\bctr}{\mathbf{CTR}}
\newcommand{\CVR}{\text{CVR}}
\newcommand{\CTR}{\text{CTR}}

\newcommand{\R}{\mathbb{R}}

\newcommand{\ba}{\mathbf{a}}

\newcommand{\bvectormu}{\boldsymbol{\mu}}
\newcommand{\bvectorx}{\mathbf{x}}

\newcommand{\diag}{\text{diag}}
\newcommand{\un}{\| \mathbf{u} \|_2}

\title{Robust autobidding for noisy conversion prediction models}

\newcommand{\shorttitle}{{\small Robust autobidding for noisy conversion prediction models}}

\author{Andrey Pudovikov\thanks{Corresponding author}\\
MSU IAI, Moscow, Russia \\
\url{a.pudovikov@iai.msu.ru}\\
\And 
Alexandra Khirianova \\
MSU IAI \\
Moscow, Russia
\And 
Ekaterina Solodneva\\
MSU IAI \\
Moscow, Russia\\
\And
Gleb Molodtsov\\
MBZUAI\\
Abu Dhabi, UAE
\And
Aleksandr Katrutsa\\
MSU IAI \\
Moscow, Russia\\
\And
Yuriy Dorn \\
IAI MSU \\
Moscow, Russia\\
\And
Egor Samosvat\\
Independent researcher\\
Moscow, Russia
}

\begin{document}
\maketitle


\begin{abstract}
Managing millions of digital auctions is an essential task for modern advertising auction systems. 
The main approach to managing digital auctions is an autobidding approach, which depends on the Click-Through Rate and Conversion Rate values. 
While these quantities are estimated with ML models, their prediction uncertainty directly impacts advertisers' revenue and bidding strategies.
To address this issue, we propose \texttt{RobustBid}, an efficient method for robust autobidding taking into account uncertainty in CTR and CVR predictions.
Our approach leverages advanced, robust optimization techniques to prevent large errors in bids if the estimates of CTR/CVR are perturbed.
We derive the analytical solution of the stated robust optimization problem, which leads to the runtime efficiency of the \texttt{RobustBid} method.
The synthetic, iPinYou, and BAT benchmarks are used in our experimental evaluation of \texttt{RobustBid}.
We compare our method with the non-robust baseline and the RiskBid algorithm in terms of total conversion volume (TCV) and average cost-per-click ($CPC_{avg}$) performance metrics.
The experiments demonstrate that \texttt{RobustBid} provides bids that yield larger TCV and smaller $CPC_{avg}$ than competitors in the case of large perturbations in CTR/CVR predictions.
\end{abstract}


\section{Introduction}
On commercial electronic platforms, a list of relevant display content, including paid options, is generated to best meet the user's needs while also maximizing revenue. 
This content is typically selected and placed through online auctions~\cite{balseiro2021landscape}, where bids are generated by automated algorithms. 
These algorithms operate within advertiser-specified constraints, analyzing past bidding data, assessing previous auction performance, and predicting future ad performance for specific requests. 
Central to this process are the tasks of the Click-Through Rate (CTR) prediction~\cite{robinson2007internet,richardson2007predicting} and the Conversion Rate (CVR) prediction~\cite{CVR1, shtoff2023improving}. 
CTR measures the likelihood of a user clicking on an ad, which is economically vital for advertisers, as it gauges the potential for a purchase to occur. 
In turn, CVR refers to the probability of conversion following a click.

The design of an auction can significantly influence bidders' behavior. 
Current advertising systems typically employ cost-per-impression, cost-per-click, or cost-per-conversion~\cite{costperimpression,costpermille} as the payment rule, which constitutes the core of the auction mechanism. 
Advertisers subsequently focus on their own metrics to estimate the effectiveness of their advertising campaigns. 
A classic approach involves maximizing the number of ad impressions, clicks, or conversions. 
However, while bidders cannot directly influence the auction design, they can strategically select their objectives based on these performance metrics. 
This strategic adaptation highlights that payment rules and metrics are inherently context-dependent; their efficacy and suitability must be evaluated in relation to the specific goals and challenges of each auction problem.

The main contributions of this work are the following.
\begin{enumerate}
    \item We propose the robust optimization problem to find the bid formula treating the uncertainty in CTR/CVR estimations. 
    \item Based on the analytical solution of the stated robust optimization problem, we suggest the \texttt{RobustBid} method, which provides proper bids even in the case of high uncertainty. 
    \item The extensive experimental evaluation of the \texttt{RobustBid} algorithm on synthetic and industrial datasets demonstrates improvement in conversion rates and reduction in the cost per click compared to the baselines.
\end{enumerate}

\section*{Related works}
Autobidding algorithms automatically determine bid prices to achieve advertisers' marketing objectives, adapting to market changes and growing data complexity~\cite{aggarwal2019autobidding, balseiro2021landscape, deng2024efficiency, aggarwal2024auto}.
These systems enable advertisers to optimize spending within budget and cost-per-click constraints~\cite{yuan2013real, pid2019, chen2024optimization, cai2017real}.

We adopt the problem formulation from~\cite{pid2019}, which considers online bidding where bids are determined sequentially for each auction without complete market information.
This formulation is widely used in the autobidding literature~\cite{aggarwal2024auto, aggarwal2019autobidding, yue2021unified, li2025gas}, making our approach broadly applicable. 
In this framework, optimal bids depend on predicted CTR and CVR values from ML models.
However, these predictions contain uncertainty that can significantly impact algorithm performance, which has received limited attention in existing works.

\paragraph{Robustness in bidding.}
Bidding algorithms are affected by the uncertainty from multiple sources.
For example, strategic uncertainty occurs when advertisers intentionally underbid to force the platform to decrease the reserve price.
The lower the reserve price is, the lower the bid that could win the auction.
\cite{balseiro2021robust, golrezaei2019dynamic, kumar2024strategically} develop robust algorithms that prevent strategic manipulation and maximize platform revenue.
However, these approaches target platform revenue maximization, while our work focuses on advertiser bidding optimization.
At the same time, uncertainty in opportunity values and win rates is considered in~\cite{qu2024double}, where robust optimization techniques are used.
Similarly, uncertainty in competing bid distributions is discussed in~\cite{kasberger2024robust}, where the min-max approach is proposed.
The latter two studies require the estimation of the ads opportunity values, whereas our approach does not. 

\paragraph{Robustness of CTR prediction.}
Since autobidding algorithms typically use the CTR estimation to provide bids~\cite{aggarwal2024auto}, uncertainty quantification of the predicted CTR values is crucial for proper bidding.   
\cite{shih2023robust} propose a clustering model, which groups bid requests with similar predicted CTRs and uses reinforcement learning to train a model for bid generation. 
This clustering model shows empirical improvements but lacks robustness guarantees. 
Study~\cite{zhang2017managing} introduces a risk management framework that incorporates prediction uncertainty via modeling bid proportional to the weighted sum of CTR and its standard deviation.
This approach has theoretical foundations but requires tuning of weights.
At the same time, work~\cite{bandi2014optimal} introduces robustness for valuation vectors by using a central limit theorem and induced confidence intervals.
However, their analysis was restricted to the offline setup, which limits its applicability to real-world systems.
To the best of our knowledge, our approach is the first to handle uncertainties in both CTR and CVR values.
We summarize the comparison of our work with the previous studies in Table~\ref{table:comparison}.

\begin{table}[!ht]
  \caption{Comparison of the proposed approach with existing alternatives. We highlight that our approach supports uncertainties in multiple sources, includes an uncertainty set, and is evaluated on real-world datasets.}
  \label{table:comparison}
  \centering
  \begin{tabular}{ccccc}
    \toprule
    \textbf{Reference} & \textbf{Multiple uncertainty sources} & \textbf{Uncertainty set}  & \textbf{Real-world data} \\
    \midrule
    \cite{pid2019} & 
      \xmark & 
      \xmark & 

      \cmark \\
    \addlinespace
    \cite{bandi2014optimal} & 
      \xmark & 
      \cmark & 

      \xmark \\
        \addlinespace
    \cite{zhang2017managing} & 
      \xmark & 
      \cmark & 

      \cmark \\
    \addlinespace
    \textbf{This paper} & 
      \cmark & 
      \cmark & 

      \cmark \\
    \bottomrule
  \end{tabular}
\end{table}

\section{Problem statement} \label{sec:problem_formulation}
This section presents the offline non-robust and robust optimization problems for identifying bids in cases where the CTR/CVR values are exact and perturbed, respectively.

\subsection{Non-robust maximization of expected number of conversions}

Consider a sequence of $T$ auctions involving $I$ items, where each item participates in every auction with corresponding bids denoted as $bid_t^i$. 
In this formulation, we assume that each advertiser has only one item, i.e., these terms are equivalent.

This bid depends on CTR value $CTR_t^i$ and conversion value $CVR^i_t$ for $i \in [1,I]$ and $t \in [1,T]$.
This dependence for a particular item $i$ is modeled through solving a specific maximization problem introduced below.
The objective in this maximization problem is the expected number of conversions for the $i$-th item, which is estimated as
\begin{equation}
     \sum_{t=1}^T x_t^i \cdot CTR_t^i \cdot CVR^i_t
    \label{eq::cnv}
\end{equation}
where $x_i^t = \begin{cases}
    1, &  \text{if item } i \text{ wins auction } t;\\
    0, & \text{otherwise.}
\end{cases}$. 
Upon winning an auction~$t$ by submitting the highest bid, the advertiser pays a winning price $wp_t$, which depends on the auction type, e.g., first-price auction, second-price auction, or others~\cite{roughgarden2010algorithmic}.
At the same time, the $i$-th advertiser has a pre-defined budget $B^i$, which induces the natural constraint of the form
\begin{equation}
    \sum_{t=1}^T x_t^i \cdot wp_t \leq B^i,
\label{eq::budget_constr}
\end{equation}
where the left-hand side corresponds to the expected spends from the wins of auctions.
In addition, the $i$-th advertiser can bound the total cost per click by a constant $C^i$, which leads to the following CPC constraint:
\begin{equation}
\frac{\sum_{t=1}^T x_t^i \cdot wp_t}{\sum_{t=1}^T x_t^i \cdot CTR_t^i} \leq C^i,
\label{eq::cpc_constr}
\end{equation}
where the left-hand side indicates the expected average cost per click.
In~(\ref{eq::cpc_constr}), we estimate expected number of clicks as $\sum_{t=1}^T x_t^i \cdot CTR_t^i$.
Thus, maximization of the objective function~(\ref{eq::cnv}) subject to constraints~(\ref{eq::budget_constr}) and~(\ref{eq::cpc_constr}) with binary variable $x_i^t$ is an integer linear programming problem, which is hard to solve in a large-dimensional case~\cite{pid2019}.
To approximate the solution, studies~\cite{pid2019,aggarwal2024auto} propose relax variable $x_i^t$ such that $x_i^t \in [0, 1]$.
Therefore, the resulting non-robust maximization of the expected number of conversions problem is stated as
\begin{equation}
    \begin{split}
        &\max _{0 \leq x_{t}^{i} \leq 1} \quad \sum _{t=1}^T x_{t}^{i} \cdot CTR_{t}^{i} \cdot CVR_{t}^{i} \\
\text {s.t. } & \quad \sum _{t=1}^T x_{t}^{i} \cdot w p_{t}^{i} \leq B^{i} \\
& \frac{\sum_{t=1}^T x_{t}^{i} \cdot w p_{t}^{i}}{\sum_{t=1}^T x_{t}^{i} \cdot C T R_{t}^{i}} \leq C^{i}. \\
    \end{split}
\label{eq::nonrobust_problem}
\end{equation}

In problem~(\ref{eq::nonrobust_problem}), we assume that the winning price is known and corresponds to the used auction type.
The main requirement for the auction is that the auction winner sets the highest bid.
As a consequence, the highest bid is the upper bound for the winning price $wp_t^i$.
If the auction mechanism satisfies this requirement, then it can be used within our framework.

To derive the dependence of bid $bid_t^i$ on CTR/CVR, \cite{pid2019} consider dual problem to problem~(\ref{eq::nonrobust_problem}) and obtains the following equation: 
\begin{equation}
    bid_t^i = \frac{1}{p+q} CVR_t^i \cdot CTR_t^i + \frac{q}{p + q} C^i \cdot CTR_t^i,
    \label{eq::non_robust_bid}
\end{equation}
where $p, q > 0$ are optimal dual variables corresponding to constraints~(\ref{eq::budget_constr}) and~(\ref{eq::cpc_constr}), respectively. 
See more details in Appendix~\ref{sec::appendix-nonrobust}. 
Equation~(\ref{eq::non_robust_bid}) demonstrates that perturbation in CTR/CVR values significantly affects the bid value.
Since the precise values of $CTR_i^t$ and $CVR_i^t$ are not known in advance in practice, they are typically estimated by machine learning models~\cite{yang2022click}.
These models estimate $CTR_i^t$ and $CVR_i^t$ with some uncertainty.
Therefore, a natural question arises about how to make the bid values robust to the uncertainty in the $CTR_i^t$ and $CVR_i^t$ values.
The following section proposes the robust modification of problem~(\ref{eq::nonrobust_problem}) based on a robust optimization framework.

\subsection{Robust conversion maximization problem from CTR/CVR uncertainty}
\label{sec:2.2}

Accurate estimations of CTR and CVR are crucial for deriving a proper solution of~(\ref{eq::nonrobust_problem}) since even a small perturbation of these values can significantly degrade the performance of the resulting bid~(\ref{eq::non_robust_bid}). 
To mitigate such effects, we apply the robust optimization framework~\cite{book_robustopt} to the autobidding problem.
The robust optimization framework requires introducing an uncertainty set $\mathcal{U}$ that corresponds to the feasible perturbations of the CTR/CVR values.
For simplicity, we consider CTR values further, while they can be replaced by CVR, and our approach processes them correctly. 
Since we assume that CTR values are estimated by a pre-trained ML model $f$, we use the corresponding loss function $L$ in the definition of $\mathcal{U}$.
We assume that to fit the model $f$, one uses train and validation datasets consisting of ground-truth CTR, denoted as $CTR$.

In particular, the loss function $L$ measures for the $i$-th advertiser how close the predicted CTR $\widehat{CTR_t^i}$ is to the ground-truth CTR ${\bctr}^i = [{CTR}_t^i]$, where $i=1,\ldots,I$ and $t=1,\ldots, T$.
Therefore, we can define an uncertainty set for the $i$-th advertiser as follows:
\begin{equation}
    \mathcal{U}^i = \left\{ \mathbf{a} = [a_t]_{t=1}^T, a_t \in [0,1] \mid L(\widehat{\bctr}^i, \mathbf{a})  \leq \varepsilon \right\},
    \label{eq::uncertainty_set}
\end{equation}
where $L(\widehat{\bctr}^i, \mathbf{a}) = \sum_{t=1}^T L (\widehat{CTR}_t^i, a_t)$, $\mathbf{a}$ is a vector of feasible CTR values and $\varepsilon > 0$ is a predefined threshold.
We assume that $\bctr^i \in \mathcal{U}^i$, $i=1,\ldots,I$ for some~$\varepsilon_0$.
According to~\cite{bertsimas2025data}, if $\varepsilon_0$ is empirical $(1+ \frac{1}{I})(1-q)$-quantile of $L(\widehat{\bctr}^i_{val}, \bctr^i )$, where $q \in (0, 1)$, then with probability  at least~$q$ the following inequality holds:
\begin{equation}
    L(\widehat{\bctr}^i_{val}, \bctr^i ) \leq \varepsilon_0.
    \label{eq::epsilon_def}
\end{equation}
To compute $\varepsilon_0$, we use $\widehat{\bctr}^i_{val}$, which is the CTR values estimated by the pre-trained model $f$ on the validation dataset. 
While $\varepsilon_0$ is computed on the validation dataset, we will use it in the general problem~(\ref{eq::nonrobust_problem}), assuming data homogeneity.

Since the loss function $L$ and uncertainty set~\eqref{eq::uncertainty_set} are independent of each item $i$, the item index is further omitted to simplify expressions. 
Following the robust optimization framework, we state the  robust modification of (\ref{eq::nonrobust_problem}):
\begin{equation}
\label{opt:robust_optimization}
    \begin{split}
        &\max_{0 \leq x_{t} \leq 1} \min_{\mathbf{a} \in \mathcal{U}} \sum_{t=1}^T x_{t} \cdot a_{t} \cdot CVR_{t} \\
\text { s.t. } & \sum_{t=1}^T x_{t} \cdot w p_{t} \leq B \\
& \sum_{t=1}^T x_{t} \cdot w p_{t} \leq C  \sum_{t=1}^T x_{t} \cdot a_{t} \\
    \end{split}
\end{equation}

The main difference between the non-robust problem~(\ref{eq::nonrobust_problem}) and the robust problem~(\ref{opt:robust_optimization}) is the objective function and uncertainty set for the $CTR$ prediction.
Note that this worst-case robust autobidding problem~(\ref{opt:robust_optimization}) depends on the choice of loss function. 
The following section presents the analytical solution of problem~(\ref{opt:robust_optimization}) and generalizes it to the case of joint uncertainty in CTR and CVR.

\section{\texttt{RobustBid} method}
\label{sec::main_robustbid}

This section presents the \texttt{RobustBid} method, which is based on the analytical solution of the robust optimization problem~(\ref{opt:robust_optimization}) for a particular uncertainty set.
The explicit form of uncertainty set $\mathcal{U}$ depends on the choice of the loss function $L$, see~(\ref{eq::uncertainty_set}). 
The straightforward loss function for the CTR prediction problem is the cross-entropy loss; however, this loss leads to an analytically intractable optimization problem.
To make the problem analytically tractable and still estimate the perturbation of the predicted CTR values $\widehat{\bctr}$, we select MSE loss to construct the uncertainty set
\begin{equation}
    \mathcal{U}_{mse} = \left\{ \mathbf{a} = [a_t]_{t=1}^T, a_t \in [0,1] \mid \frac12 \left \| \widehat{\bctr} - \mathbf{a} \right \|_2^2  \leq \varepsilon_a \right\}.
    \label{eq::uncertainty_set_mse}
\end{equation}
Incorporating the uncertainty set $\mathcal{U}_{mse}$ in problem~(\ref{opt:robust_optimization}) makes it possible to solve this problem analytically.
If instead of uncertainty in CTR estimates one considers uncertainty in CVR estimates, then the derived formulas can be used in such a case, see details in Appendix~\ref{sec::appendix_only_cvr}.
Moreover, the \texttt{RobustBid} method supports uncertainty in \emph{both} CTR and CVR estimates.
The analytical solution to the corresponding robust optimization problem is presented in this section as well. 
Thus, the uncertainty set $\mathcal{U}_{mse}$ provides a reasonable trade-off between the tractability of the resulting robust optimization problem and the correctness of the uncertainty measurements.   

\subsection{Bidding formulas for individual CTR/CVR uncertainties} 
\label{sec:mse_main}

This section presents the analytical bidding formulas in the case of uncertainty CTR and CVR values separately.
We derive the bidding formula for the case of uncertainty in CTR from analytical solution of problem~(\ref{opt:robust_optimization}), where $\mathcal{U} = \mathcal{U}_{mse}$. 
To solve this problem, we introduce a slack variable $s$ such that the original problem is rewritten in the following form:
\begin{equation}
    \begin{split}
        & \max_{0 \leq x_t \leq 1, s} s \\ 
        \text{s.t. } & \sum _{t=1}^T x_{t} \cdot w p_{t} \leq B, \quad \sum _{t=1}^T x_{t} \cdot w p_{t} \leq C  \sum _{t=1}^T x_{t} \cdot a_{t} \\
& s \leq \min_{\mathbf{a} \in \mathcal{U}_{mse}}  \sum_{t=1}^T x_{t} \cdot a_{t} \cdot CVR_{t}
    \end{split}
    \label{opt::slack_robust}
\end{equation}

The minimization problem in the last constraint assumes that $x_t$ and $CVR_t$ are known and only the vector $\mathbf{a}$ is unknown.
Therefore, this problem has a linear objective function in the form $\mathbf{m}^\top \mathbf{a}$, where $m_t = x_t \cdot CVR_t$, and quadratic inequality constraint $\frac12 \|  \widehat{\bctr} - \mathbf{a} \|_2^2 \leq \varepsilon_a$.
We eliminate inequality constraints of the form $0 \leq a_t \leq 1$ to make this problem analytically tractable and further discuss the cases in which they will be satisfied automatically.
Since the resulting problem is convex, we use KKT optimality conditions and derive the optimal solution $\tilde{\mathbf{a}}$ such that:
\begin{equation}
    \tilde{\ba} = \widehat{\bctr}-\frac{\mathbf{m}}{\|\mathbf{m}\|_{2}} \alpha , \quad \alpha = \sqrt{ 2 \varepsilon_a}.
    \label{eq::ctr_worst_mse}
\end{equation}
The detailed derivation of~(\ref{eq::ctr_worst_mse}) is presented in Appendix~\ref{sec:internal_opt}. 
Note that $\tilde{a_t} \leq 1$ since $CTR_t \leq 1$ and $\alpha, m_t \geq 0$. 
To satisfy constraint $\tilde a \geq 0$ we set $\varepsilon_a$ such that 
\begin{equation}
    \varepsilon_a \leq \varepsilon_m = \min_t \frac{\widehat{CTR_t} \cdot \| \mathbf{CVR}\|_2}{CVR_t}.
\label{eq::def_varepsilon_a}
\end{equation}
The estimated $\varepsilon_a = \varepsilon_m$ can lead to excessive pessimistic probabilistic guarantees according to~(\ref{eq::epsilon_def}). 

After that, we substitute $\tilde{\ba}$ in problem~(\ref{opt::slack_robust}) and denote $\alpha=\sqrt{2 \varepsilon_a}$ to get the following optimization problem with linear objective function and convex quadratic constraints:
\begin{equation}
\begin{split}
&\max_{\mathbf{x}, \mathbf{y}, s} s \\
\text { s.t. } &  y_{t}=x_{t} \cdot C V R_{t} \quad \forall t \\
& s \leq \mathbf{y}^{\top} \widehat{\bctr} - \alpha\|\mathbf{y}\|_{2} \\
&\sum_{t=1}^T x_{t} \cdot w p_{t} \leq B \\
&\sum_{t=1}^T x_{t} \cdot w p_{t} \leq C \cdot [\mathbf{x}^{\top} \widehat{\bctr} - \alpha\|\mathbf{x}\|_{2}  ] \\
& 0 \leq x_{t} \leq 1, \forall t\\
\end{split}
\end{equation}

Note that quadratic constraints from this maximization problem, including the 2-norm of vectors $\mathbf{x}$ and $\mathbf{y}$ can be expressed as cone constraints.
In particular, denote by $K_2 = \{ (\mathbf{u}, v) \mid ||\mathbf{u}||_2 \leq  v\}$ the cone induced by 2-norm, then we have the following equivalence:
\[ 
{\alpha}||\mathbf{y}||_2 \leq -s + \mathbf{y}^\top \widehat{\bctr} \Leftrightarrow \begin{pmatrix}
    \alpha \mathbf{ y} \\ -s +  \mathbf{y}^\top \widehat{\bctr}
\end{pmatrix} \in K_2
\]
\[
C \alpha\|\mathbf{x}\|_{2}\leq \mathbf{x}^\top (C \cdot \widehat{\bctr} -  \mathbf{wp})   \Leftrightarrow \begin{pmatrix}
        C\alpha\mathbf{x} \\  \mathbf{x}^\top (C \cdot \widehat{\bctr} -  \mathbf{wp}) 
    \end{pmatrix} \in K_2,
\]
where $\mathbf{wp} = [wp_1, \ldots, wp_T]$ is a vector of winning prices.
The resulting cone maximization problem can be solved analytically with KKT optimality conditions. 
Moreover, the dual problem can also be derived and solved in closed form.
Thus, we can compute the optimal dual variables $p^{*}_{ctr}$ and $q^*_{ctr}$ corresponding to budget and CPC constraints.
Similar to the approach from~\cite{pid2019}, we use $p^{*}_{ctr}$, $q^*_{ctr}$ to construct the bid formula: 
\begin{equation}
    \label{eq::robust_bid_ctr}
    bid_t = \frac{1}{p^*_{ctr}+q^*_{ctr}} CVR_t \cdot \widehat{CTR_t} + \frac{q^*_{ctr}}{p^*_{ctr} + q^*_{ctr}} C \cdot \widehat{CTR_t} + \delta^{ctr}_t,
\end{equation}
where  $\delta^{ctr}_t = \begin{cases}
    - \frac{\alpha}{p^*_{ctr} + q^*_{ctr}} \left( \frac{Cq^*_{ctr}}{\sqrt{|\mathcal{T}|}} + \frac{CVR_t^2}{\sqrt{\sum_{l \in \mathcal{T}} CVR_l^2}} \right) & t \in \mathcal{T}, \\
    0 & t \notin \mathcal{T}, \end{cases}$ is the additive perturbation to~(\ref{eq::non_robust_bid}) and a set of timestamps $\mathcal{T}$ is the following
    \begin{equation}
        \mathcal{T} = \{t \mid CTR_t \cdot CVR_t  + q^*_{ctr} \cdot C \cdot CTR_t \leq (p^*_{ctr}+q^*_{ctr}) wp_t  \}.
    \label{inactive_index_formula}
    \end{equation}

If only uncertainty in CVR is assumed, then the corresponding uncertainty set $\mathcal{V}_{mse}$ can be defined as
\begin{equation}
    \label{eq::cvr_mse_unsertainty_set}
    \mathcal{V}_{mse} = \left\{ \mathbf{a} = [a_t]_{t=1}^T, a_t \in [0,1] \mid \frac12 \left \| \widehat{\mathbf{CVR}} - \mathbf{a} \right \|_2^2  \leq \varepsilon_b \right\}.
\end{equation}
In this case, the robust version of problem~(\ref{eq::nonrobust_problem}) becomes easier to solve since CVR uncertainty affects only the objective function and does not modify constraints.
The corresponding dual problem also has an analytical solution, which includes optimal dual variables $p^*_{cvr}$ and $q^*_{cvr}$.
Following the same steps as in the derivation of~(\ref{eq::non_robust_bid}) and~(\ref{eq::robust_bid_ctr}), these dual variables can be used to construct the bid formula robust to CVR perturbation:
\begin{equation}
\label{eq::main_bid_cvr}
bid_t = \frac{1}{p^*_{cvr}+q^*_{cvr}} \widehat{CVR_t} \cdot CTR_t + \frac{q^*_{cvr}}{p^*_{cvr} + q^*_{cvr}} C \cdot CTR_t+ \delta_t^{cvr},
\end{equation}
where $\delta_t^{cvr} = \begin{cases}
     - \frac {\alpha'}{p^*_{cvr}+q^*_{cvr}} \frac{ CTR_t}{ \sqrt{\sum_{l \in \mathcal{T}} CTR^2_l}},  & t \in \mathcal{T}\\
     0, & t \notin \mathcal{T}.
\end{cases}$, $\alpha' = \sqrt{2\varepsilon_b}$ and $\mathcal{T}$ is defined in~ (\ref{inactive_index_formula}).
The detailed derivaton of~(\ref{eq::main_bid_cvr}) is presented in Appendix~\ref{sec::appendix_only_cvr}.
Thus, the final bid formula for CVR uncertainties~(\ref{eq::main_bid_cvr}) looks similar to the bid formula for CTR uncertainties~(\ref{eq::robust_bid_ctr}) except for the form of the perturbation item, where the CVR is replaced with CTR and the term related to the CPC constraint is dropped.

\subsection{Bidding formula for joint CTR and CVR uncertainties} 
\label{sec:joint_mse_main}

This section extends a robust optimization approach to address uncertainties in \emph{both} CTR and CVR predictions simultaneously. 
This setup represents a more comprehensive modeling of uncertainties in the bidding process.
Since we have introduced uncertainty sets for CTR $\mathcal{U}_{mse}$ and for CVR $\mathcal{V}_{mse}$ in Section~\ref{sec:mse_main}, we can state the robust optimization problem which treats the Cartesian product of these uncertainty sets:
\begin{equation}
\label{opt:joint_mse_problem}
\begin{split}
    & \max_{0 \leq x_t\leq 1} \min_{(\mathbf{a}, \mathbf{b}) \in \mathcal{U}_{mse} \times \mathcal{V}_{mse}}
    \sum_{t=1}^T x_t\,a_t\,b_t; \\
    \text{s.t. } & 
    \sum_{t=1}^T x_t\,wp_t\le B \\
    & \frac{\sum_{t=1}^T x_t\,wp_t}{\sum_{t=1}^T x_t\,a_t}\le C.
\end{split}
\end{equation}

This problem is more complex than the single uncertainty case, as it involves the interaction between uncertainties in both CTR and CVR values. 
To solve problem~(\ref{opt:joint_mse_problem}), we use advanced techniques from convex optimization theory, including the S-lemma and Schur complement theorem.
Below, we provide the main steps to derive a robust optimal bid formula.

To simplify notations, we use denote $a^0_t=\widehat{\CTR_t}, b^0_t=\widehat{\CVR_t}$, then introduce perturbations \(\delta a_t=a_t-a^0_t,\;\delta b_t=b_t-b^0_t\) and obtain new form for inequality in uncertainty set 
\begin{equation}
    \|\mathbf{\delta a}\|_2^2 \le r_a^2=2\varepsilon_a, \|\mathbf{\delta b}\|_2^2\le r_b^2=2\varepsilon_b.
\label{eq::uncertan_set_as_perturb}
\end{equation}
The objective is rewritten as follows: 
\[
  \sum_{t=1}^T x_t\,a_t\,b_t
  =\sum_t x_t\,(a^0_t+\delta a_t)(b^0_t+\delta b_t)
  =\mathbf{z}^\top \mathbf{Q}\mathbf{z} + 2\mathbf{q}^\top \mathbf{z} + c,
\]
where
\[
\mathbf{z}=\begin{pmatrix}\mathbf{\delta a}\\ 
  \mathbf{\delta b}
  \end{pmatrix},\quad
  \mathbf{D} = \diag(x_1,\dots,x_T),\quad
  \mathbf{Q} = \begin{pmatrix}0 & \tfrac12 \mathbf{D}\\
  \tfrac12 \mathbf{D} & 0\end{pmatrix},
\]
\[ \mathbf{q}=\tfrac12\begin{pmatrix}\mathbf{D}\,\mathbf{b}^0\\
\mathbf{D}\,\mathbf{a}^0
  \end{pmatrix},\quad
  c=\sum_{t=1}^T x_t\,a^0_t\,b^0_t.
\]
Then we introduce a slack scalar variable \(s\) similar to Section~\ref{sec:mse_main}:
\[
s \leq \min_{\|\mathbf{\delta a}\|_2 \le r_a,\|\mathbf{\delta b}\|_2\le r_b}
(\mathbf{z}^\top \mathbf{Q} \mathbf{z} +2\mathbf{q}^\top \mathbf{z} + c)
\]
which is equivalent to
\[
\mathbf{z}^\top \mathbf{Q} \mathbf{z} + 2\mathbf{q}^\top \mathbf{z} + (c-s) \ge 0
\quad \forall \mathbf{z}.
\]
To reformulate constraints~(\ref{eq::uncertan_set_as_perturb}) in the quadratic forms with positive semi-definite matrices, we use the S-lemma.
For this purpose, we rewrite inequalities:
\[
g_1(\mathbf{z})\equiv r_a^2-\|\mathbf{\delta a}\|^2_2 = r_a^2 - \mathbf{\delta a}^T \mathbf{I}\mathbf{\delta a}\;\ge0,
\]
\[
g_2(\mathbf{z})\equiv r_b^2-\|\mathbf{\delta b}\|^2_2= r_b^2 - \mathbf{\delta b}^T \mathbf{I} \mathbf{\delta b} \ge 0
\]
and apply the inhomogeneous S‑lemma \cite{book_robustopt}. 
Then, 
\begin{equation}
    g_1(\mathbf{z})\ge0,\;g_2(\mathbf{z})\ge0
\;\Rightarrow\;
\mathbf{z}^\top \mathbf{Q} \mathbf{z} +2\mathbf{q}^\top \mathbf{z} + (c-s)\ge0
\quad\forall \mathbf{z}
\label{eq::s_lemma}
\end{equation}
holds if and only if there exist \(\lambda_a,\lambda_b\ge0\) such that
\[
\mathbf{M} = 
\begin{pmatrix}
\mathbf{Q} + \diag(\lambda_a \mathbf{I},\lambda_b \mathbf{I}) & \mathbf{q}\\[4pt]
\mathbf{q}^\top & (c-s-\lambda_a r_a^2-\lambda_b r_b^2)
\end{pmatrix}
\succeq0.
\]
To simplify this constraint, we apply the Schur complement theorem~\cite{book_robustopt}, so
\(\mathbf{M} \succeq 0\) is equivalent to the two convex constraints:
\[
\begin{pmatrix}
\lambda_a \mathbf{I}_T & \frac12 \mathbf{D}\\[3pt]
\frac12 \mathbf{D}    & \lambda_b \mathbf{I}_T
\end{pmatrix} \succeq 0 
\]
\[
(c - s -\lambda_a r_a^2 - \lambda_b r_b^2 ) - \mathbf{q}^\top (\mathbf{Q} + \diag(\lambda_a \mathbf{I}, \lambda_b \mathbf{I}))^{-1} \mathbf{q} \geq 0.
\]
By combining the fact that $(\mathbf{Q} + \diag(\lambda_a \mathbf{I}, \lambda_b \mathbf{I}))^{-1} \succeq 0$, since inversion preserves positive semi-definiteness, and applying the definitions of $\mathbf{Q}$ and $\mathbf{q}$, we obtain:
\[
\begin{pmatrix}
\lambda_a \mathbf{I_T} & \frac12 \mathbf{D}\\
\tfrac12 \mathbf{D}    & \lambda_b \mathbf{I_T} 
\end{pmatrix} \succeq 0 
\quad\text{and}\quad
-\lambda_a r_a^2 - \lambda_b r_b^2 + c - s
\geq0.
\]
Here, the first is a \(2T\times2T\) linear matrix inequality, and the second is an affine bound.
However, the positive semi-definite constraint can be simplified by the Schur complement theorem and the diagonal structure of $D$, we obtain:
\[
\begin{cases}
    \lambda_a \geq0,\\
    \lambda_a \lambda_b \geq \frac{1}{4} x_t^2 \quad \forall t \in 1, ..., T.
\end{cases}
\]
Therefore, the resulting simplified problem can be expressed as:

\begin{equation}
\begin{split}
\label{eq::main_joint_mse}
    & \max_{x_1, \ldots,x_T,\lambda_a,\lambda_b}\quad -\lambda_a\,r_a^2 - \lambda_b\,r_b^2 + \sum_{t=1}^T x_t\,a^0_t\,b^0_t - \sum_{t=1}^T \frac{2 x_t^2 [\lambda_b (b_t^0)^2 + \lambda_a (a_t^0)^2] - 2 x_t^3 a_t^0 b_t^0}{4 \lambda_a \lambda_b - x_t^2} \\
    \text{s.t. } & 
    \sum_{t=1}^T x_t\,wp_t \le B,\\
    & \sum_{t=1}^T x_t\,wp_t - C\sum_{t=1}^T x_t\,a^0_t + C\,r_a\,\|x\|_2 \le 0,\\
    & \lambda_a \lambda_b \geq \frac{1}{4} x_t^2,\\
    & \lambda_a \geq 0, \lambda_b \geq 0, 0 \leq x_t \leq 1
\end{split}
\end{equation}

The solution of problem~\eqref{eq::main_joint_mse} and the solution of the corresponding dual problem lead to the optimal bidding formula in the case of joint CTR and CVR uncertainties:
\begin{equation}
\label{bid_Joint_MSE_main}
   bid_t = \frac{1}{p^*_{joint}+q^*_{joint}} \widehat{CVR_t} \cdot \widehat{CTR_t} + \frac{q^*_{joint}}{p^*_{joint} + q^*_{joint}} C \cdot \widehat{CTR_t} + \delta^{joint}_t,
\end{equation}
where bid perturbation is expressed as  
\[
\delta^{joint}_t = \begin{cases}
    -\frac{\alpha}{p^*_{joint} + q^*_{joint}} \left( \frac{q^*_{joint}}{\sqrt{|\mathcal{T}|}} + A(\lambda_a^*, \lambda_b^*, \widehat{CTR_t}, \widehat{CVR_t}) \right) & t \in \mathcal{T} \\
    0 & \textstyle{t \notin \mathcal{T},}
    \end{cases}
    \]
    where $p^*_{joint}$ and $q^*_{joint}$ are optimal dual variables corresponding to the CPC and budget constraints, $\lambda_a^*, \lambda_b^*$ are optimal dual variables corresponding to the uncertainty bounds, and 
    \begin{equation}
\label{inactive_index_formula_joint} \mathcal{T} = \{t \mid CTR_t CVR_t  + C q \text{CTR}_t - A(\lambda_a, \lambda_b, \widehat{CTR_t}, \widehat{CVR_t}) \leq (p+q) wp_t \}
    \end{equation} is the set of active indices and helpful $A(\lambda_a, \lambda_b, \widehat{CTR_t}, \widehat{CVR_t})$ is defined as:
\begin{equation}
\begin{split}
    \label{helpful_a_joint}
    A(\lambda_a^*, \lambda_b^*, \widehat{CTR_t}, \widehat{CVR_t}) & = \frac{4[\lambda_a^* \widehat{CTR_t^2} + \lambda_b^* \widehat{CVR_t^2}]  - 6 \widehat{CTR_t} \widehat{CVR_t}}{(4 \lambda_a^* \lambda_b^* - 1)} +\\  
    & + \frac{4(\lambda_a^* \widehat{CTR_t^2} + \lambda_b^* \widehat{CVR_t^2} - \widehat{CTR_t} \widehat{CVR_t})}{(4 \lambda_a^* \lambda_b^* - 1)^2}
    \end{split}
\end{equation}
The detailed derivation of the presented formula can be found in Appendix~\ref{sec::appendix_joint}.

Equation~\eqref{bid_Joint_MSE_main} requires the exact values of the $p^*_{joint}, q^*_{joint}, \lambda_a^*, \lambda_b^*$, therefore the similar approach from~\cite{pid2019} is used: determine these variables on the history and apply them to the next steps.
Note that this bid formula has a similar structure to the bid formulas~\eqref{eq::robust_bid_ctr} and~\eqref{eq::main_bid_cvr} corresponding to the individual uncertainties in CTR and CVR.
However, equation~\eqref{bid_Joint_MSE_main} includes an additional term that captures the interaction between CTR and CVR uncertainties. 

\section{Numerical experiments}
\label{sec::experiments}
This section evaluates the performance of our \texttt{RobustBid} algorithm through experiments on both synthetic and real-world datasets. 
We compare our approach against two baselines: the NonRobustBid algorithm~\cite{pid2019} based on~(\ref{eq::non_robust_bid}), and the RiskBid algorithm~\cite{zhang2017managing}. 
All results are fully reproducible using the code available in our repository~\footnote{\url{https://anonymous.4open.science/r/robust-autobidding-noisy-conversions-302B}}.

\subsection{Datasets and environment design}

To simulate real-world uncertainty in CTR predictions, various levels of noise $\varepsilon^a$ and $\varepsilon^b$ are introduced into $CTR$ and $CVR$ data. The variation of $\varepsilon$ is naturally carried out in permissible normalization boundaries, upper bounds on budget $B$ and cost per click $C$ are chosen so that equation~\eqref{eq::nonrobust_problem} has a finite solution (see values of~$B$ and~$C$ in Appendix~\ref{sec::appendix-experiments}).

Advertisers have access to their historical data, including bids, winning prices, wins, accumulated clicks, and spending. 
Before placing a bid, they can also access the current values $\widehat{CTR_t^i}$ and $\widehat{CVR_t^i}$. 
The auction outcomes and click events are determined based on the ground truth values $CTR_t^{i}$ and $CVR_t^{i}$. 
Moreover, the winning prices are not allowed before the auction ends. 
Therefore, an advertiser gets winning prices from the historical data.
For the autobidding task, parameters are updated at each step by solving a constrained nonlinear optimization problem~(\ref{opt:robust_optimization}) using the advertiser's full historical data.
When a bid is placed, the advertiser is charged the minimum of the full bid amount and the remaining budget. 
The experimental evaluation was carried out using synthetic and industrial datasets.
The $\varepsilon_a, \varepsilon_b$ lie in range $[10^{-6}, 10^{-2}]$ for all experiments.

\paragraph{Synthetic dataset}
This dataset is used to evaluate performance under controlled uncertainties in CTR and CVR values.
The first-price auction mechanism is simulated over $T = 100$ auctions with $n = 10$ participating advertisers. 
The ground-truth CTR values $CTR_t^{i}$ and CVR values $CVR_t^{i}$ are randomly assigned within the range $(0.01, 0.1]$.

\paragraph{iPinYou dataset.}
The first-price auction environment consists of $n=10$ advertisers competing over $T=100$ auctions. 
For the first 9 advertisers, bid distributions are derived from the iPinYou dataset. 
Since the original dataset contains only 2-3 discrete bid values, kernel density estimation is applied to generate a continuous bid distribution that better reflects real-world behavior.
The ground-truth $CTR_t^{i}$ and $CVR_t^{i}$ values and their perturbations are sampled in the same way as in the synthetic dataset. 
The 10th advertiser follows the proposed bidding strategy with parameters adjusted to the iPinYou dataset bid statistics.

\paragraph{BAT dataset.}
The study~\cite{khirianova2025bat} proposed the new open BAT dataset for benchmarking autobidding algorithms.
The dataset consists of thousands of First-Price Auction records from real advertising campaigns. Unlike synthetic and IPinYou datasets, where auctions occur simultaneously, the BAT distributes them across time, following real-world patterns. 
The $CTR_t^{i}$ values in this dataset range from 0.0017 to 0.63, and $CVR_t^{i}$ values range from 0.001 to 0.3.

\subsection{Baselines}
\texttt{NonRobustBid}. To address the issue of non-robustness, the dual problem formulation proposed in \cite{pid2019} is adopted (see Appendix \ref{sec::appendix-nonrobust} for details). Following the approach outlined in the original paper, there are two adjusted dual variables, which correspond to the budget and CPC constraints.

The baseline solution comes directly from solving the original optimization problem exactly, outperforming parameter tuning approaches using reinforcement learning or controllers, which typically approximate the optimal solution with a predictably worse result. Thus, we compare the robust solution against the non-robust optimal solution to isolate and demonstrate the specific benefits of incorporating uncertainty handling.

\texttt{RiskBid}~\cite{zhang2017managing}. An adaptation of the \texttt{RiskBid} algorithm was used as an additional baseline. This method is designed to enhance bidding robustness by incorporating an estimate of the $CTR$ standard deviation, $CTR_{std}$, derived from historical data, into the bidding formula~\eqref{eq::non_robust_bid}: $CTR_{risk}=CTR -\alpha \cdot CTR_{std}$, where $\alpha$ is a hyperparameter.

\subsection{Metrics}
To evaluate the performance of the considered autobidding algorithms, we use the following metrics. 
\textit{Total Conversion Value} (TCV) quantifies the aggregate conversion performance across all advertising campaigns and auctions, computed as:
$$\text{TCV} =\sum _{t=1}^{T} \sum _{i=1}^{I} CTR_t^{i}\cdot CVR_t^{i}\cdot x_t^i,$$
where $CTR_t^{i}$ and $CVR_t^{i}$ represent the click-through and conversion rates respectively for campaign $i$ at time $t$, and $x_t^i$ a binary indicator of auction win.
\textit{Average Cost-Per-Click} ($CPC_{\text{avg}}$) provides the mean cost efficiency across all advertising campaigns. 
It represents the ratio of total spending to total clicks across all campaigns:    $$CPC_{\text{avg}}=\frac{\sum_{i=1}^I  \sum_{t=1}^T x_t^i\cdot bid_t^i}{\sum_{i=1}^I  \sum_{t=1}^T x_t^i\cdot CTR_t^{i}}.$$

\subsection{Individual uncertainty}
\begin{figure}[h]
    \centering
    
    \begin{minipage}[t]{0.32\linewidth}
        \centering
        Synthetic \\[5pt]
        \includegraphics[width=\textwidth]{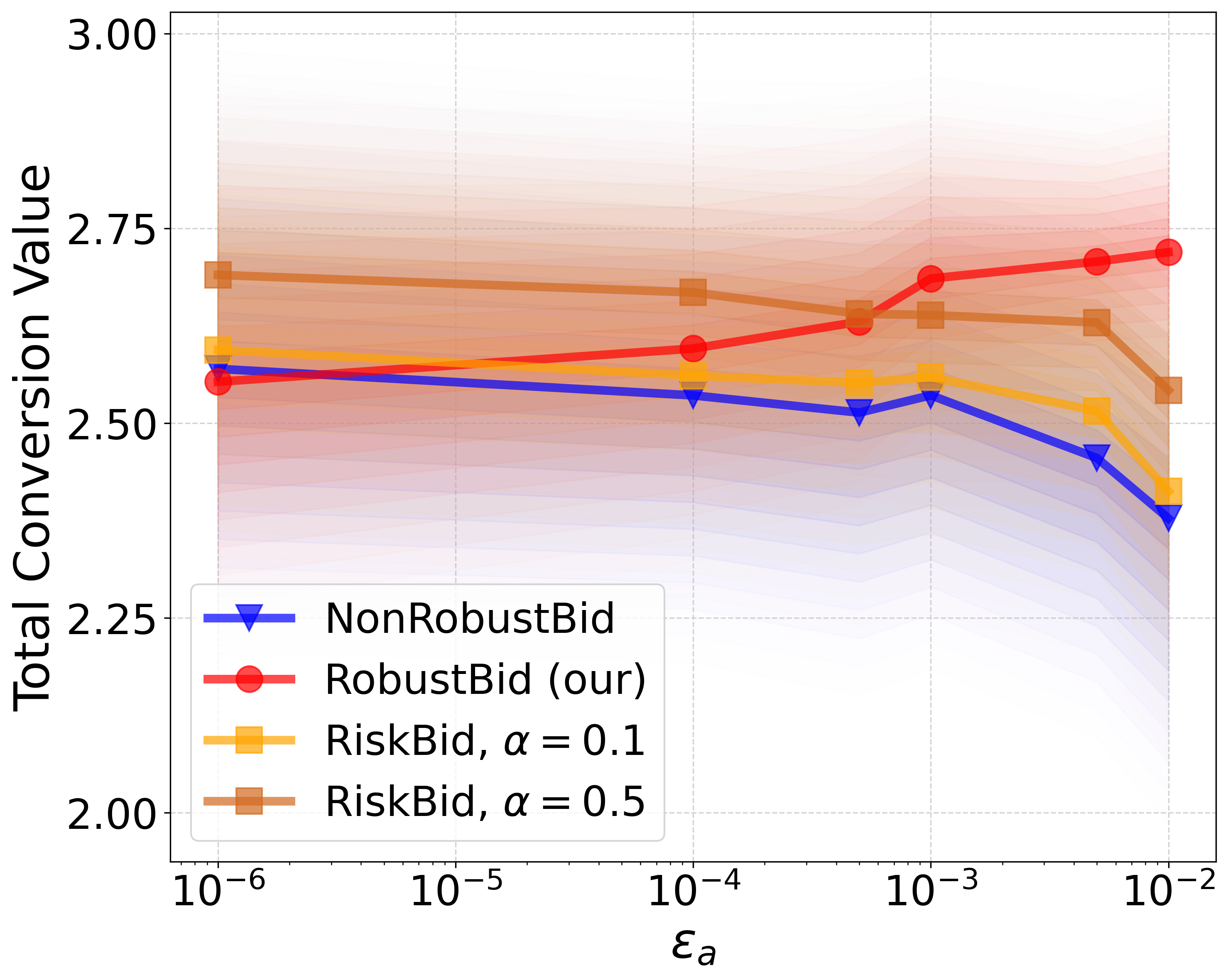}
    \end{minipage}
    \hfill
    \begin{minipage}[t]{0.32\linewidth}
        \centering
        IPinYou \\[5pt]
        \includegraphics[width=\textwidth]{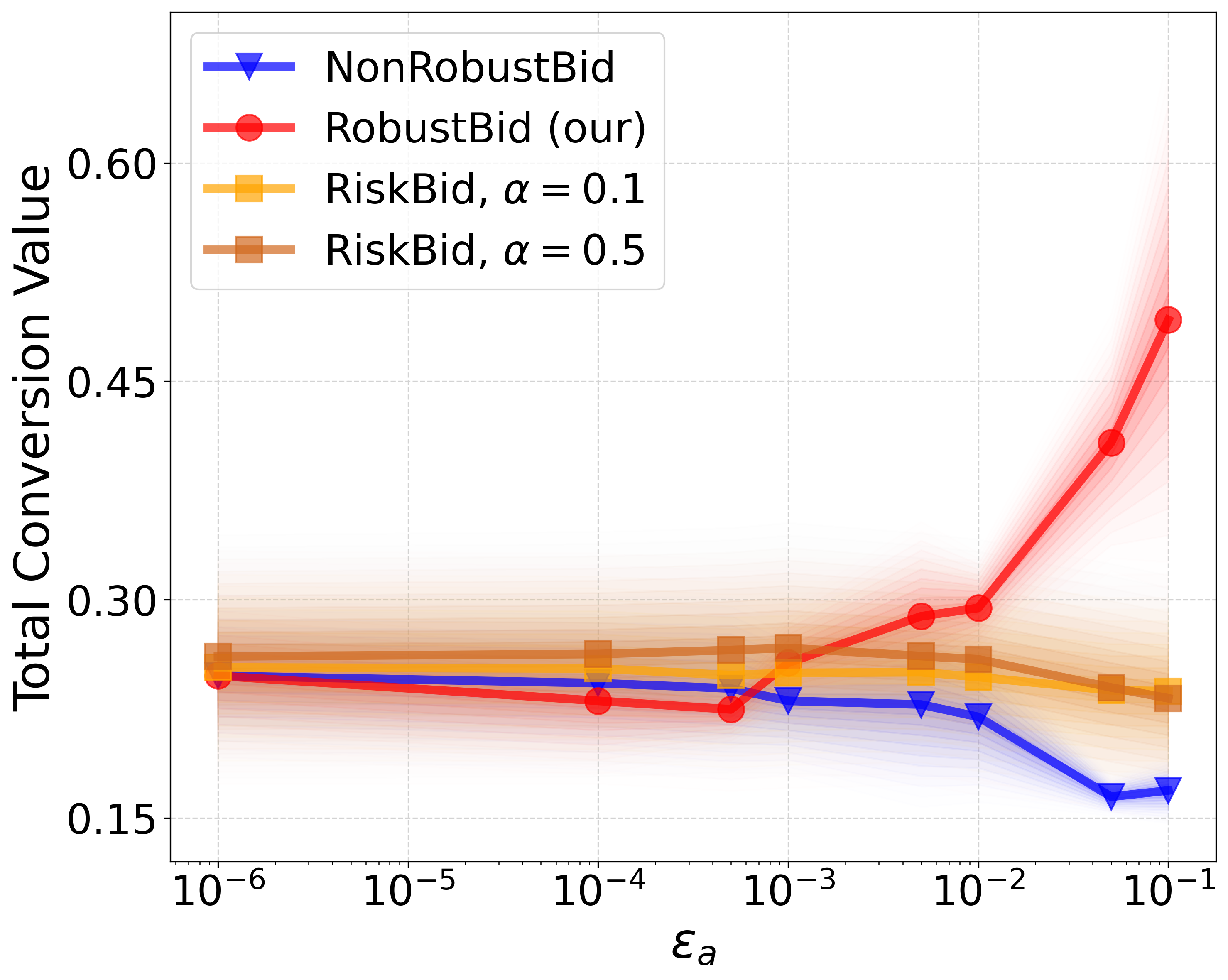}
    \end{minipage}
    \hfill
    \begin{minipage}[t]{0.32\linewidth}
        \centering
        BAT  \\[5pt]
        \includegraphics[width=1\textwidth]{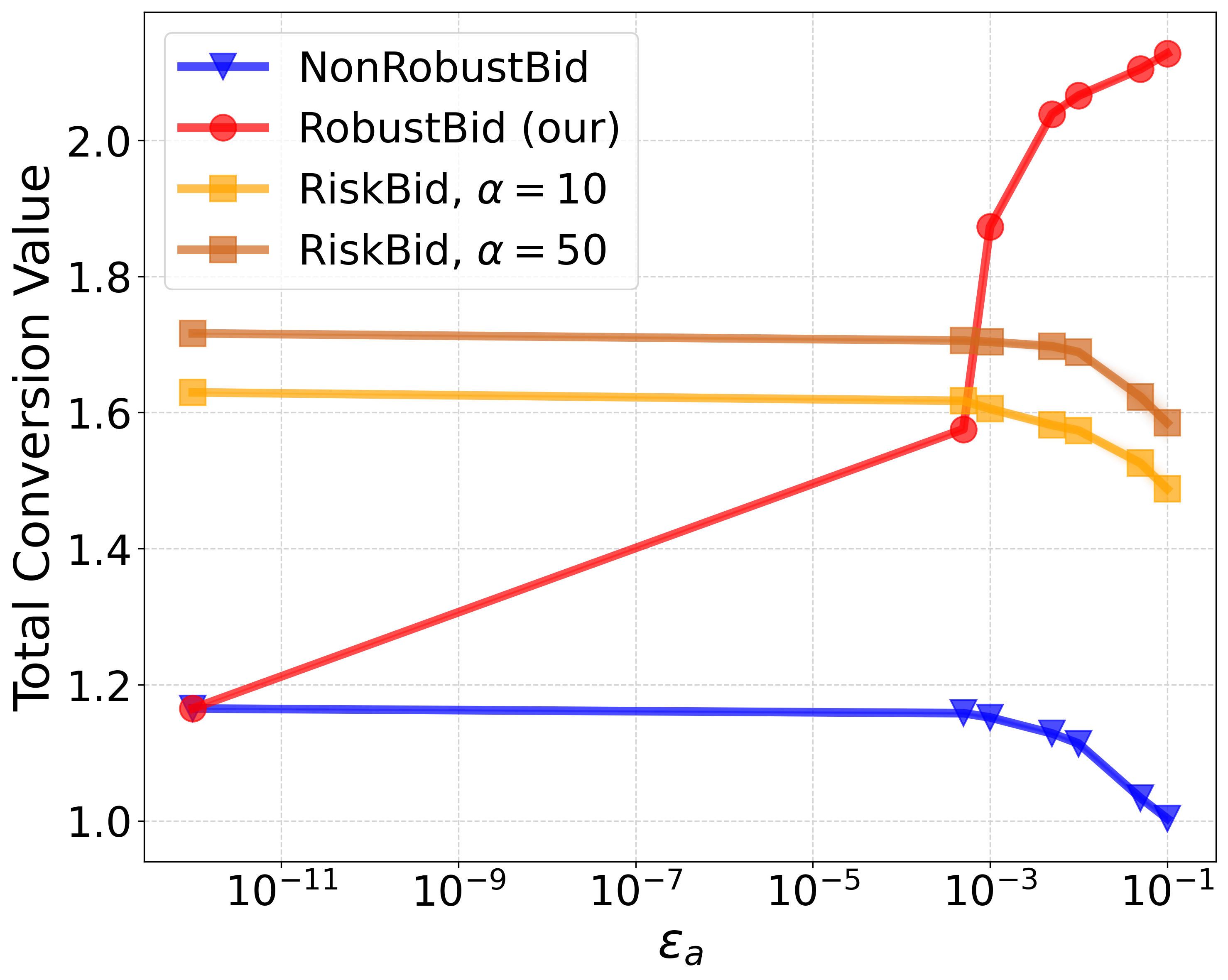}
    \end{minipage}
    \begin{minipage}[t]{0.32\linewidth}
        \centering
        \includegraphics[width=1\textwidth]{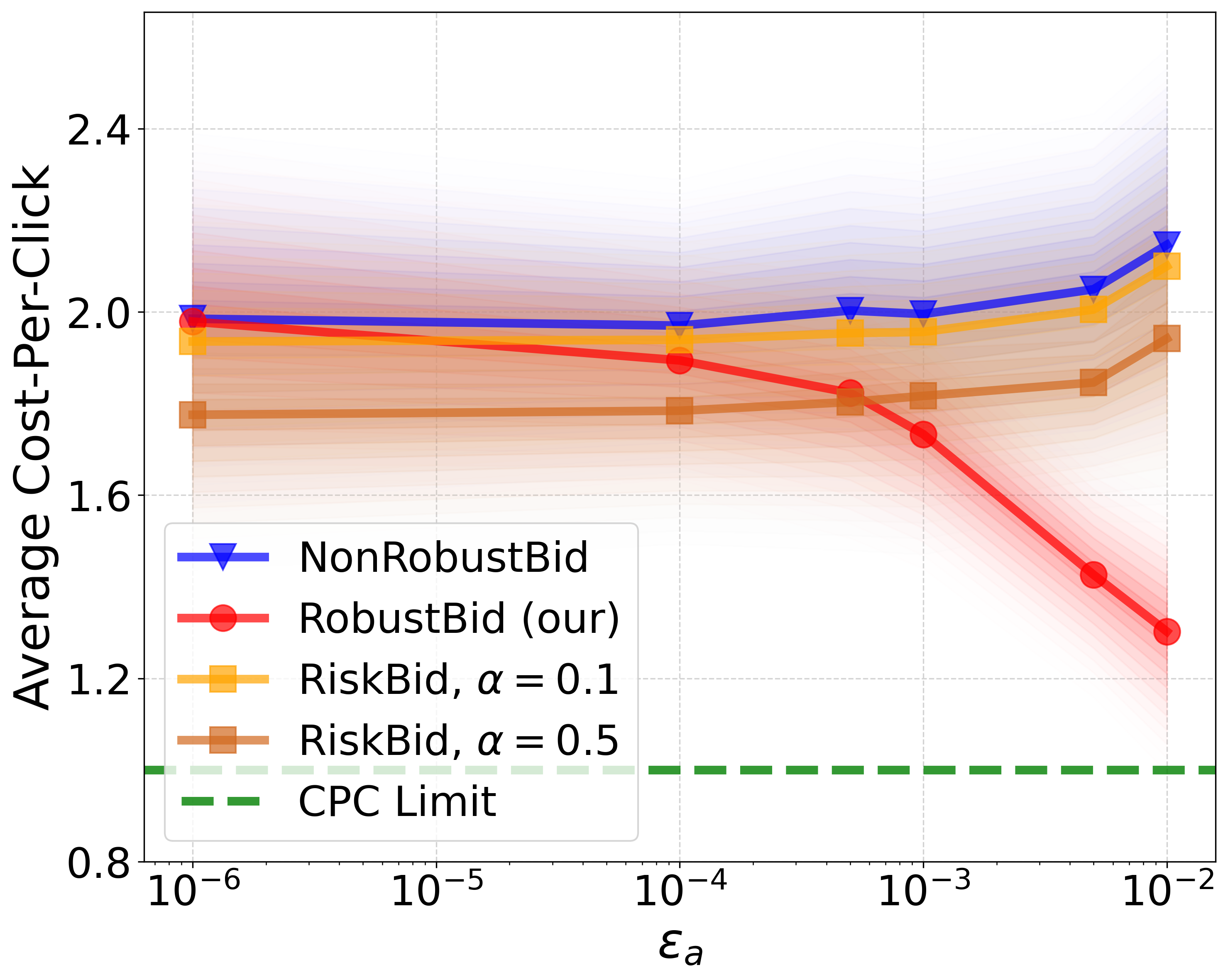}
    \end{minipage}
    \hfill
    \begin{minipage}[t]{0.32\linewidth}
        \centering
        \includegraphics[width=1\textwidth]{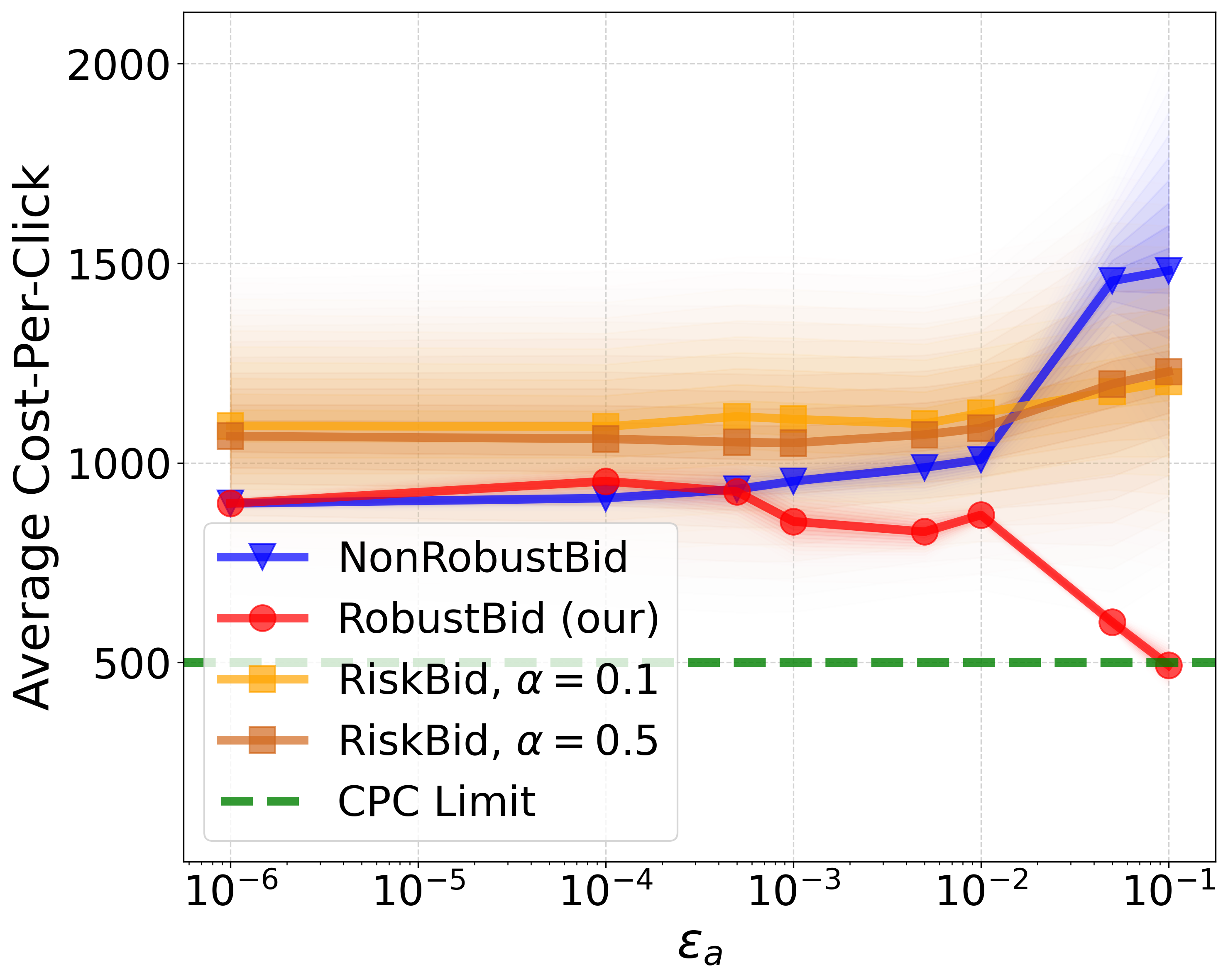}
    \end{minipage}
    \hfill
    \begin{minipage}[t]{0.32\linewidth}
        \centering
        \includegraphics[width=1\textwidth]{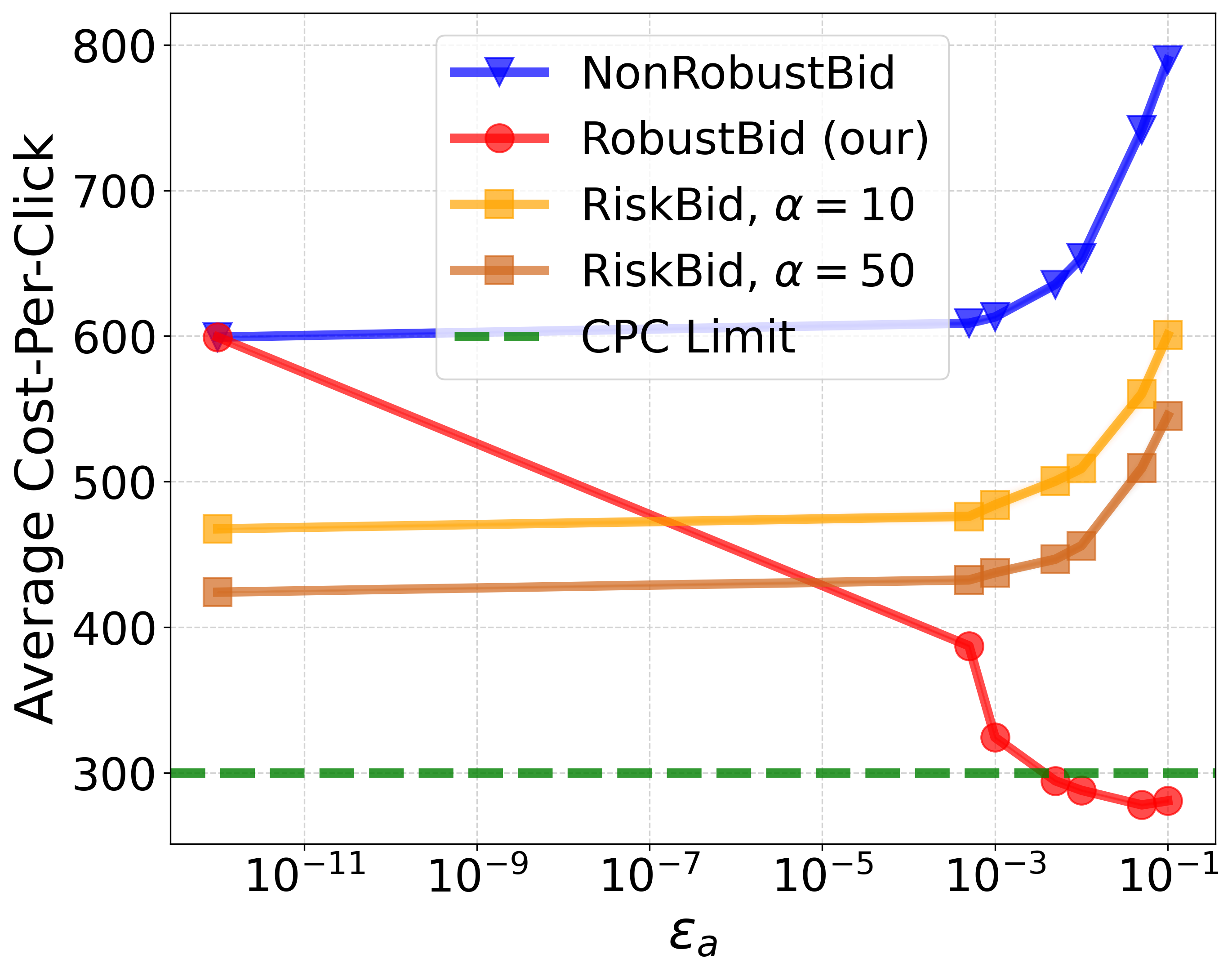}
    \end{minipage}
    \caption{Experimental results with $CTR$ uncertainty and $CVR$ accurate estimate. The green dotted line represents the upper bound on the CPC constraint. 
    Root mean squared deviation is specified in the gradient shading form. $\alpha$ is a risk-averse parameter, the larger $\alpha$ corresponds to less risky strategies.}
    \label{fig:ctr_uncertain}
\end{figure}

The dependencies of the metrics on the $CTR$ uncertainty level, assuming CVR is known exactly, are presented in Figure~\ref{fig:ctr_uncertain}. Note that with zero epsilon, the results of the \texttt{RobustBid} and \texttt{NonRobustBid} naturally converge to the same values in case. \texttt{RiskBid} performs better than \texttt{NonRobustBid}, and the \texttt{RobustBid} outperforms both baselines. The dependencies of the metrics on the $CTR$ and $CVR$ uncertainty levels are presented in Figure~\ref{fig:ctr_uncertain}.

\subsection{Joint uncertainty}

\begin{figure}[!h]
    \centering
    \begin{minipage}[b]{0.33\linewidth}
        \centering
        \includegraphics[width=1.25\textwidth]{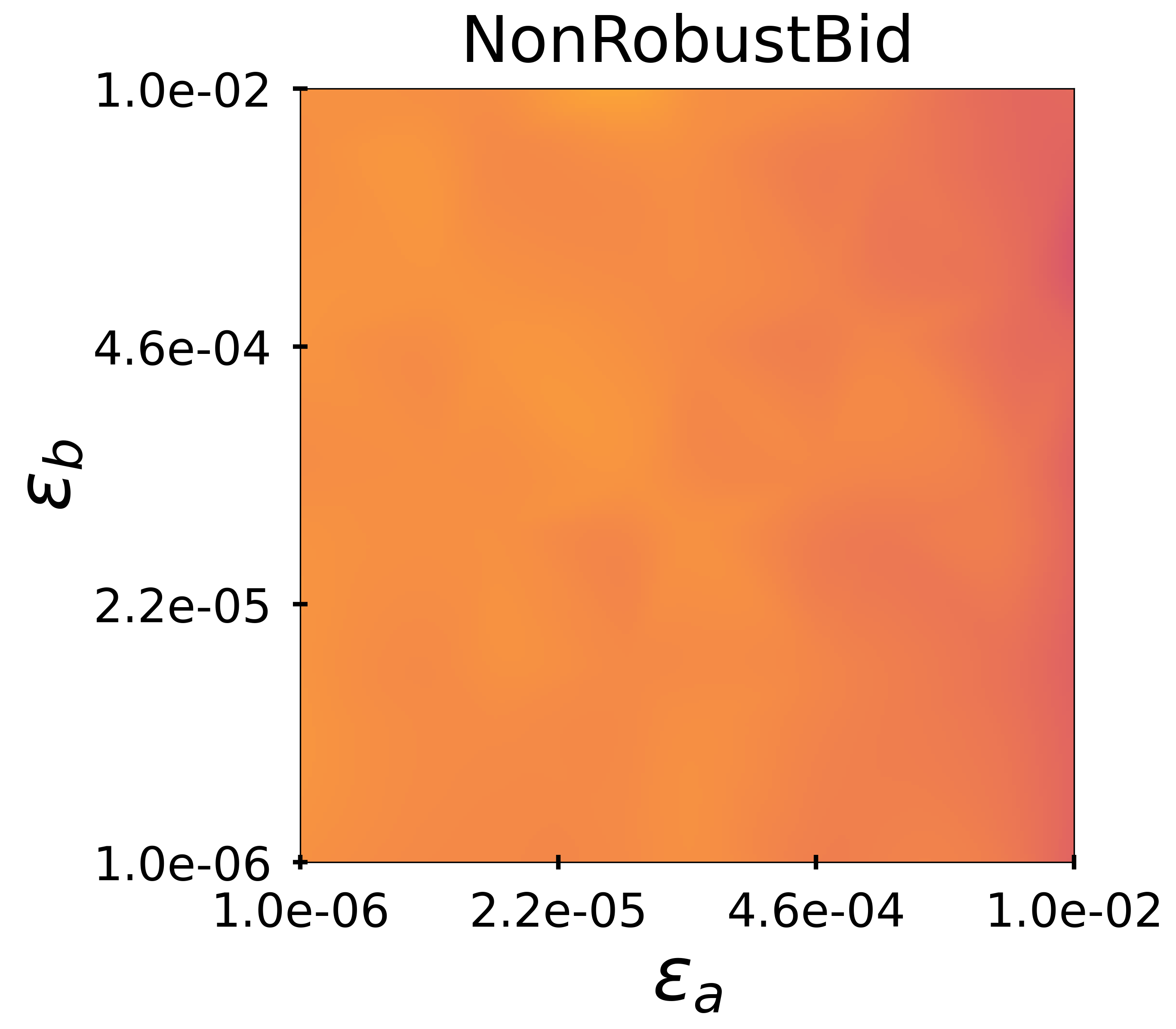}
    \end{minipage}
    \hfill
    \begin{minipage}[b]{0.33\linewidth}
        \centering
        \includegraphics[width=1.25\textwidth]{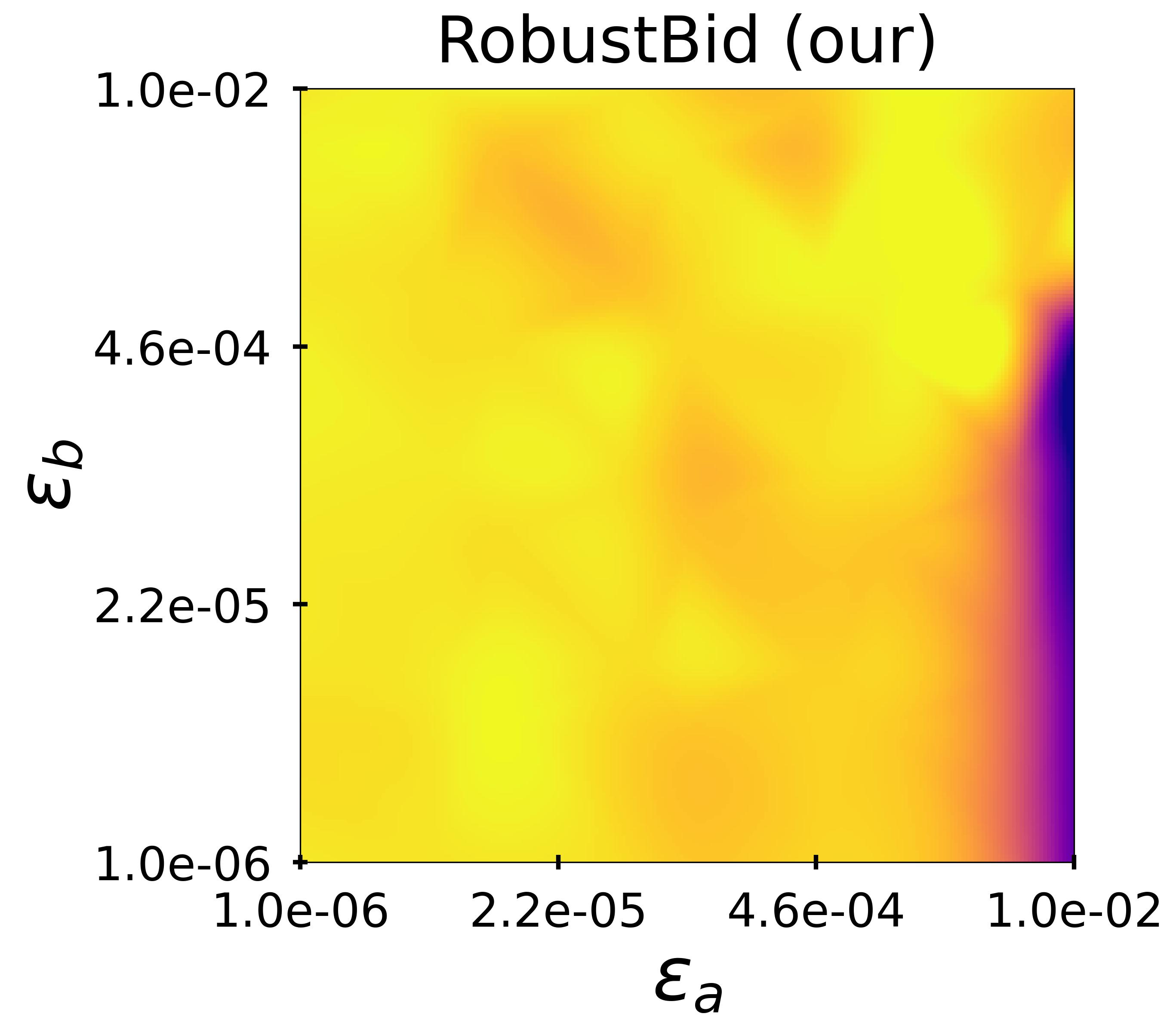}
    \end{minipage}
    \hfill
    \begin{minipage}[b]{0.15\linewidth}
        \centering
        \includegraphics[width=0.5\textwidth]{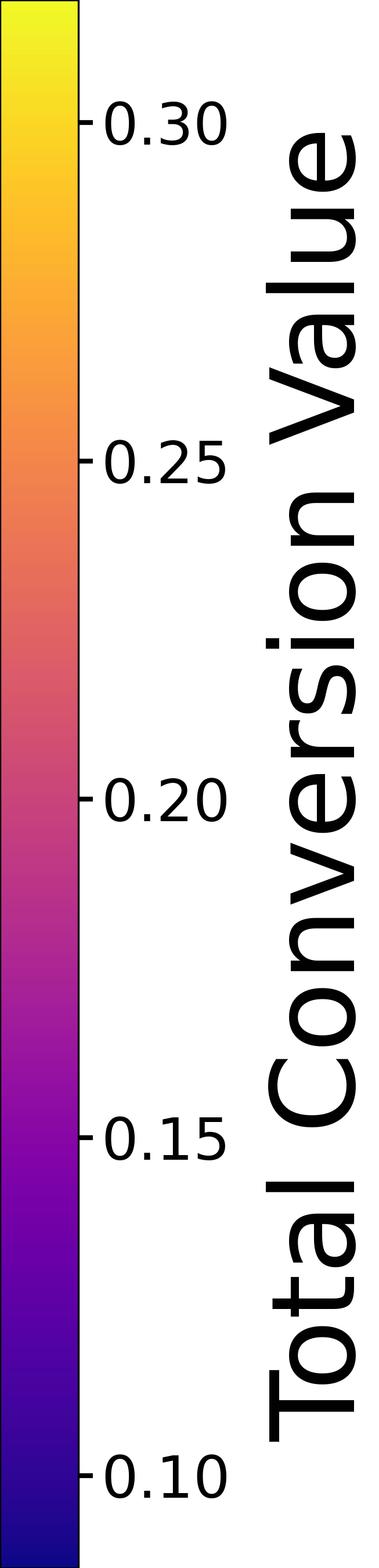}
    \end{minipage}
    
    
    \begin{minipage}[b]{0.33\linewidth}
        \centering
        \includegraphics[width=1.25\textwidth]{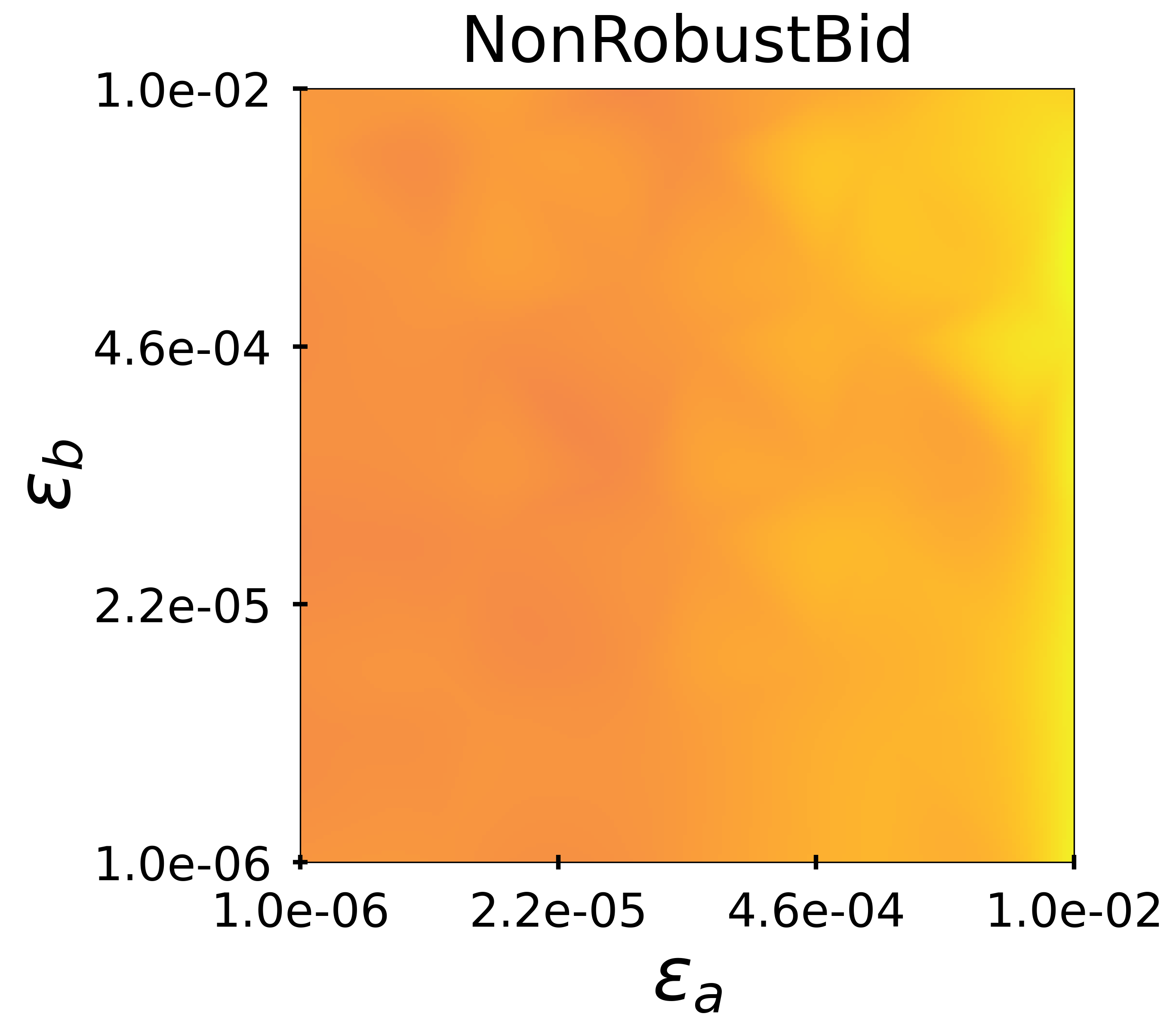}
    \end{minipage}
    \hfill
    \begin{minipage}[b]{0.33\linewidth}
        \centering
        \includegraphics[width=1.25\textwidth]{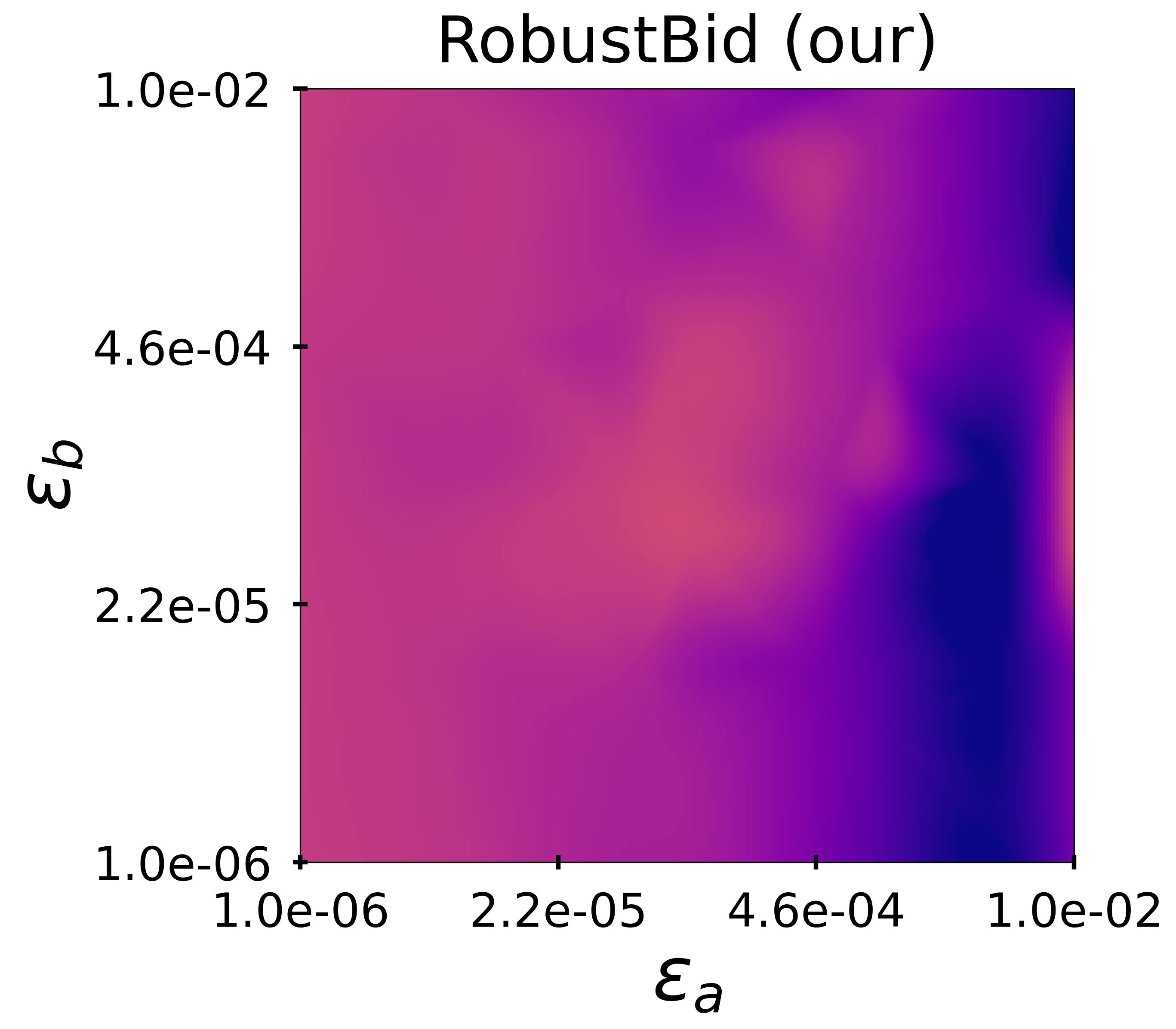}
    \end{minipage}
    \hfill
    \begin{minipage}[b]{0.15\linewidth}
        \centering
        \includegraphics[width=0.5\textwidth]{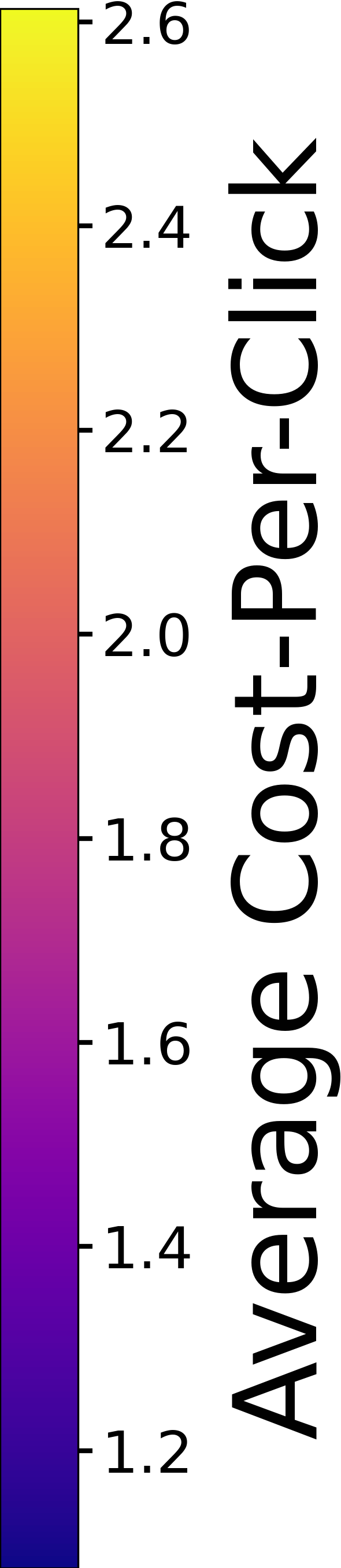}
    \end{minipage}
    
    \caption{Heatmaps with comparison $TCV$ and $CPC_{avg}$ for Synthetic dataset. 
    \texttt{NonRobustBid} metrics remain approximately the same, slightly decreasing at large values of $\varepsilon_a$ and $\varepsilon_b$. 
    \texttt{RobustBid} performs better overall, with $TCV$ decreasing only at large $\varepsilon_a$, while maintaining lower $CPC_{avg}$.}
    \label{fig::CTRCVR_synth}
\end{figure}


\begin{figure}[!h]
    \centering
    \begin{minipage}[b]{0.33\linewidth}
        \centering
        \includegraphics[width=1.25\textwidth]{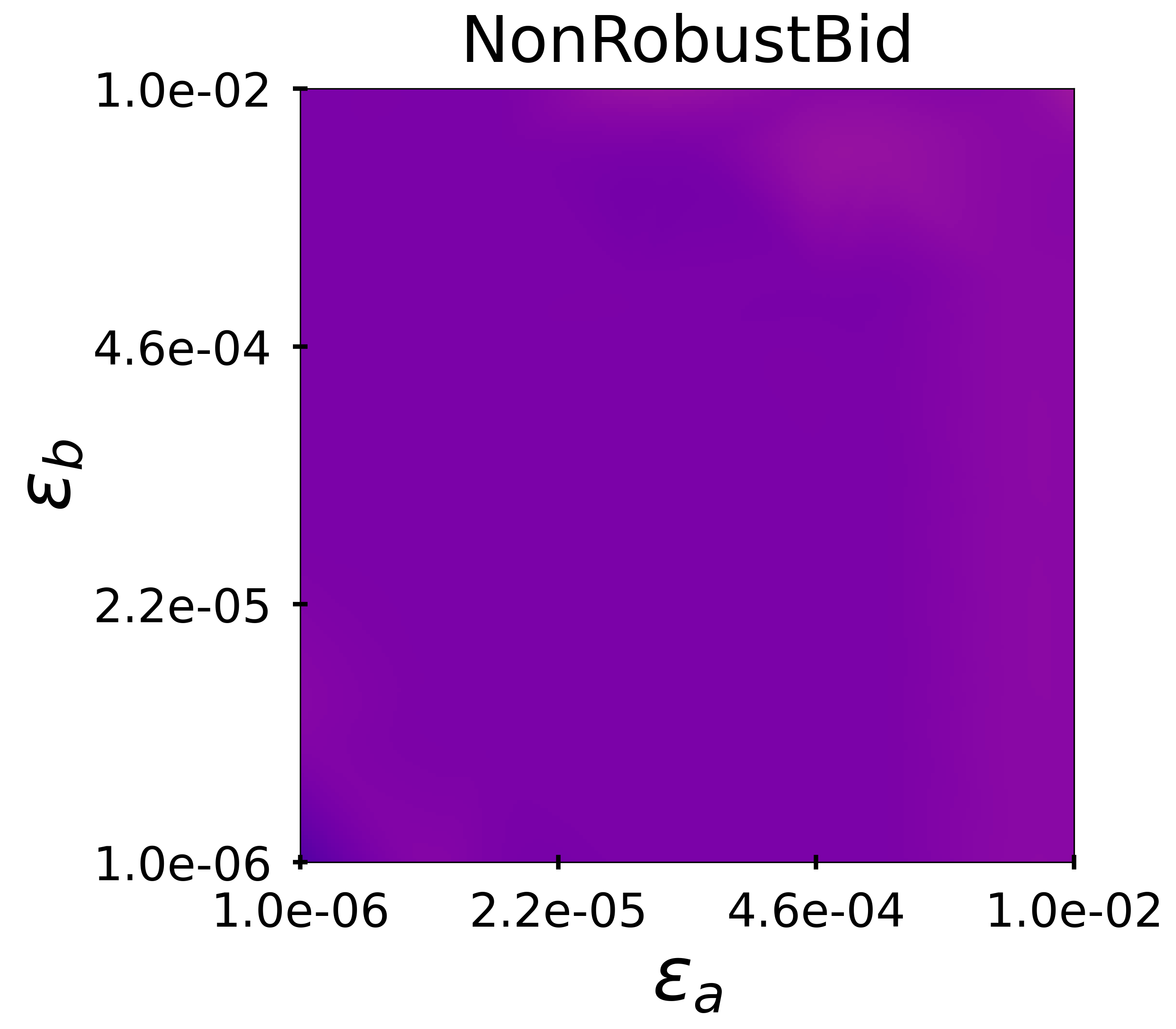}
    \end{minipage}
    \hfill
    \begin{minipage}[b]{0.33\linewidth}
        \centering
        \includegraphics[width=1.25\textwidth]{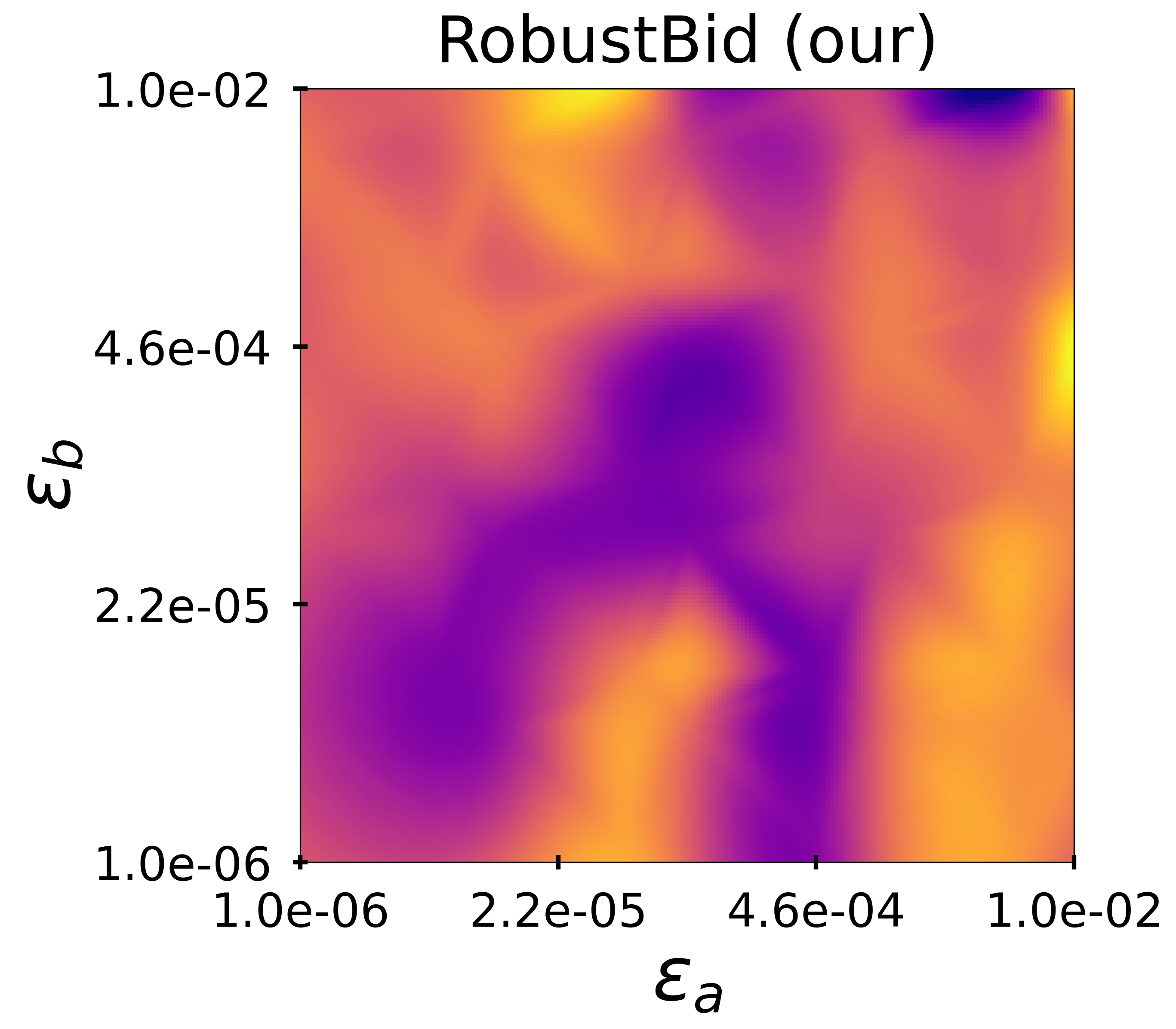}
    \end{minipage}
    \hfill
    \begin{minipage}[b]{0.15\linewidth}
        \centering
        \includegraphics[width=0.5\textwidth]{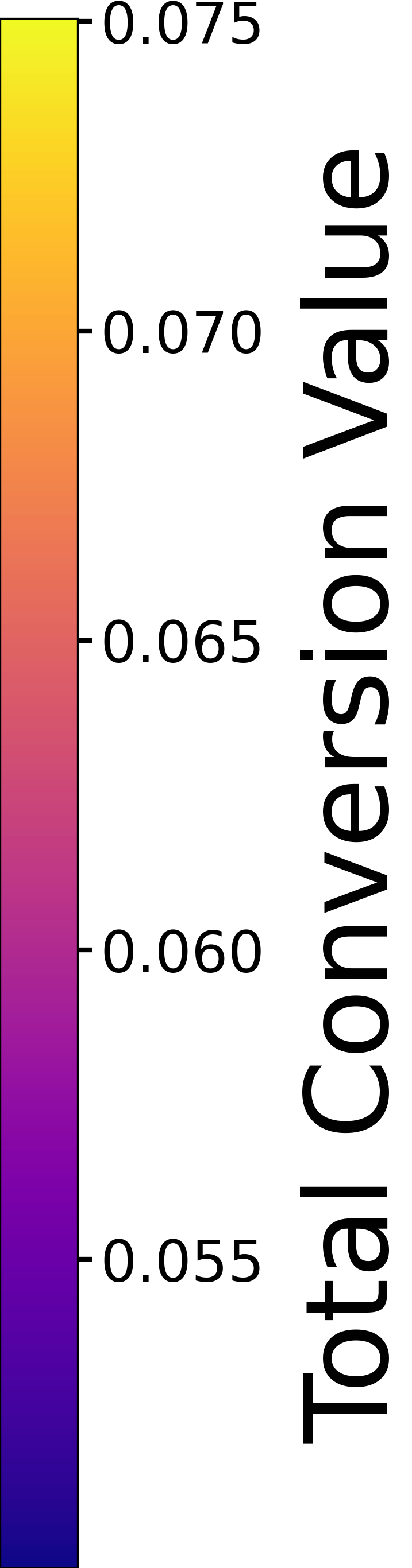}
    \end{minipage}
    
    
    \begin{minipage}[b]{0.33\linewidth}
        \centering
        \includegraphics[width=1.25\textwidth]{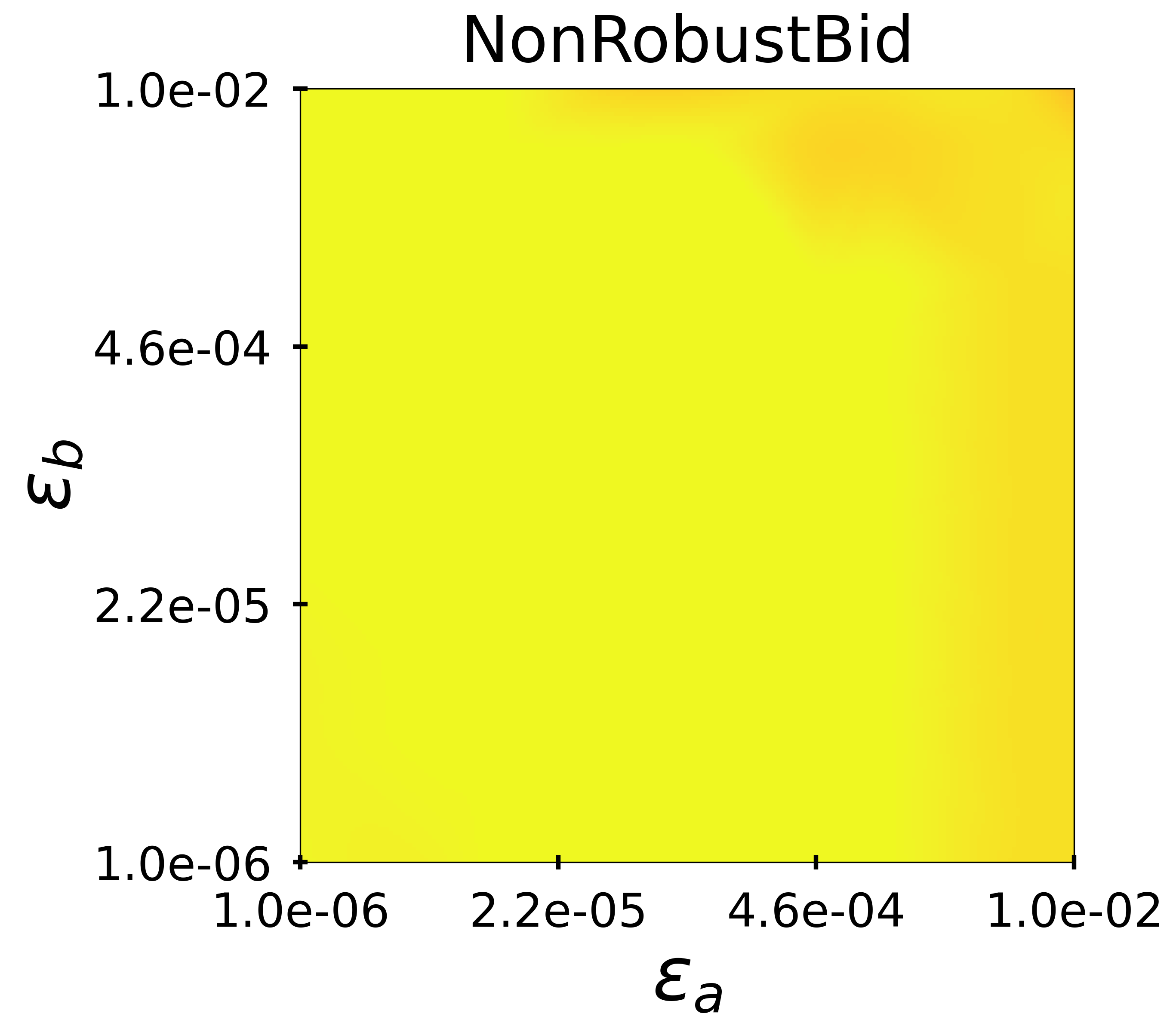}
    \end{minipage}
    \hfill
    \begin{minipage}[b]{0.33\linewidth}
        \centering
        \includegraphics[width=1.25\textwidth]{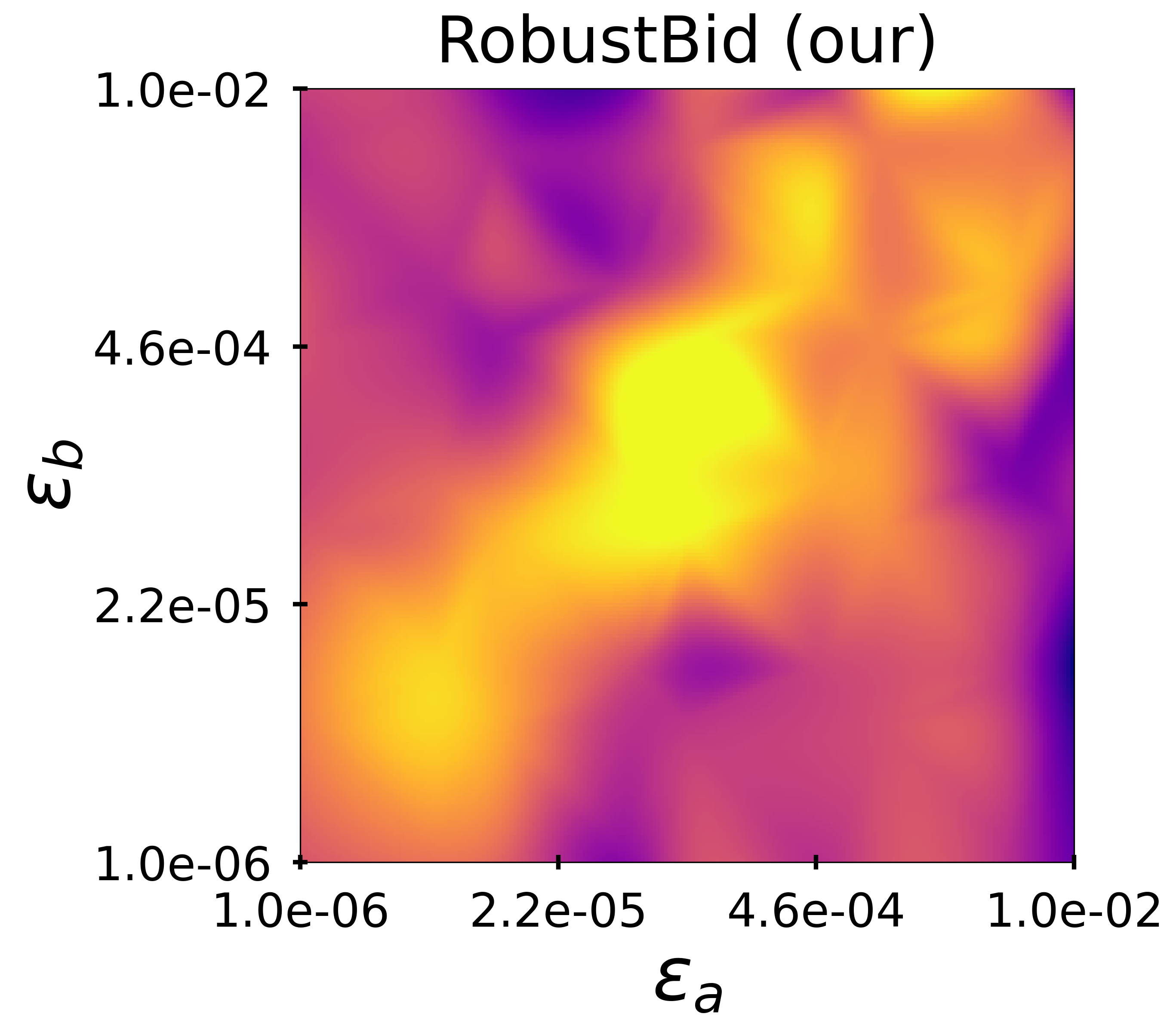}
    \end{minipage}
    \hfill
    \begin{minipage}[b]{0.15\linewidth}
        \centering
        \includegraphics[width=0.5\textwidth]{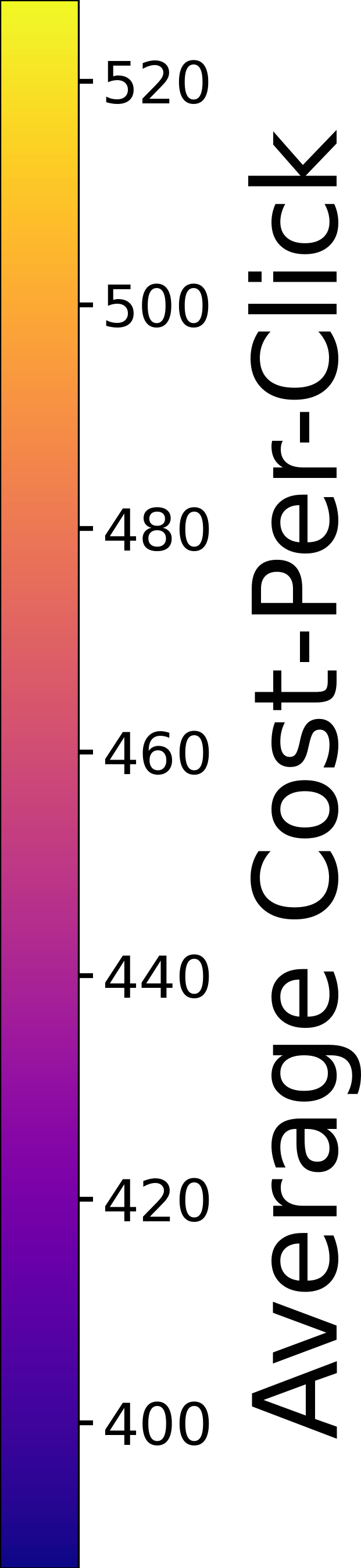}
    \end{minipage}
    
    \caption{Heatmaps with comparison $TCV$ and $CPC_{avg}$ for iPinYou dataset.
    \texttt{NonRobustBid} metrics remain approximately the same across all $\varepsilon_a$ and $\varepsilon_b$.
    In contrast, \texttt{RobustBid} metrics show significant variation while demonstrating better performance overall.
    }
    \label{fig::CTRCVR_ipinyou}
\end{figure}


\begin{figure}[!h]
    \centering
    \begin{minipage}[b]{0.33\linewidth}
        \centering
        \includegraphics[width=1.25\textwidth]{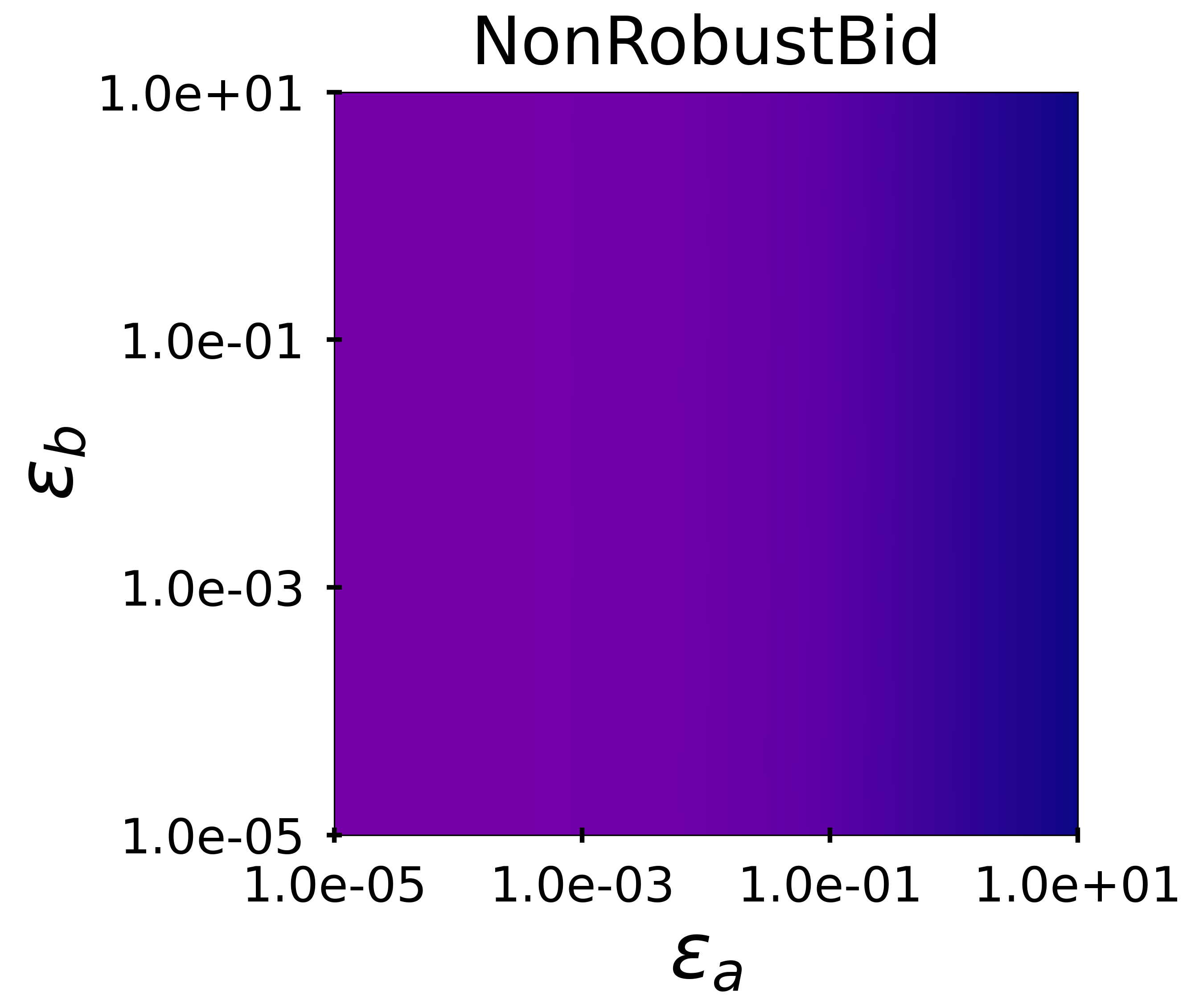}
    \end{minipage}
    \hfill
    \begin{minipage}[b]{0.33\linewidth}
        \centering
        \includegraphics[width=1.25\textwidth]{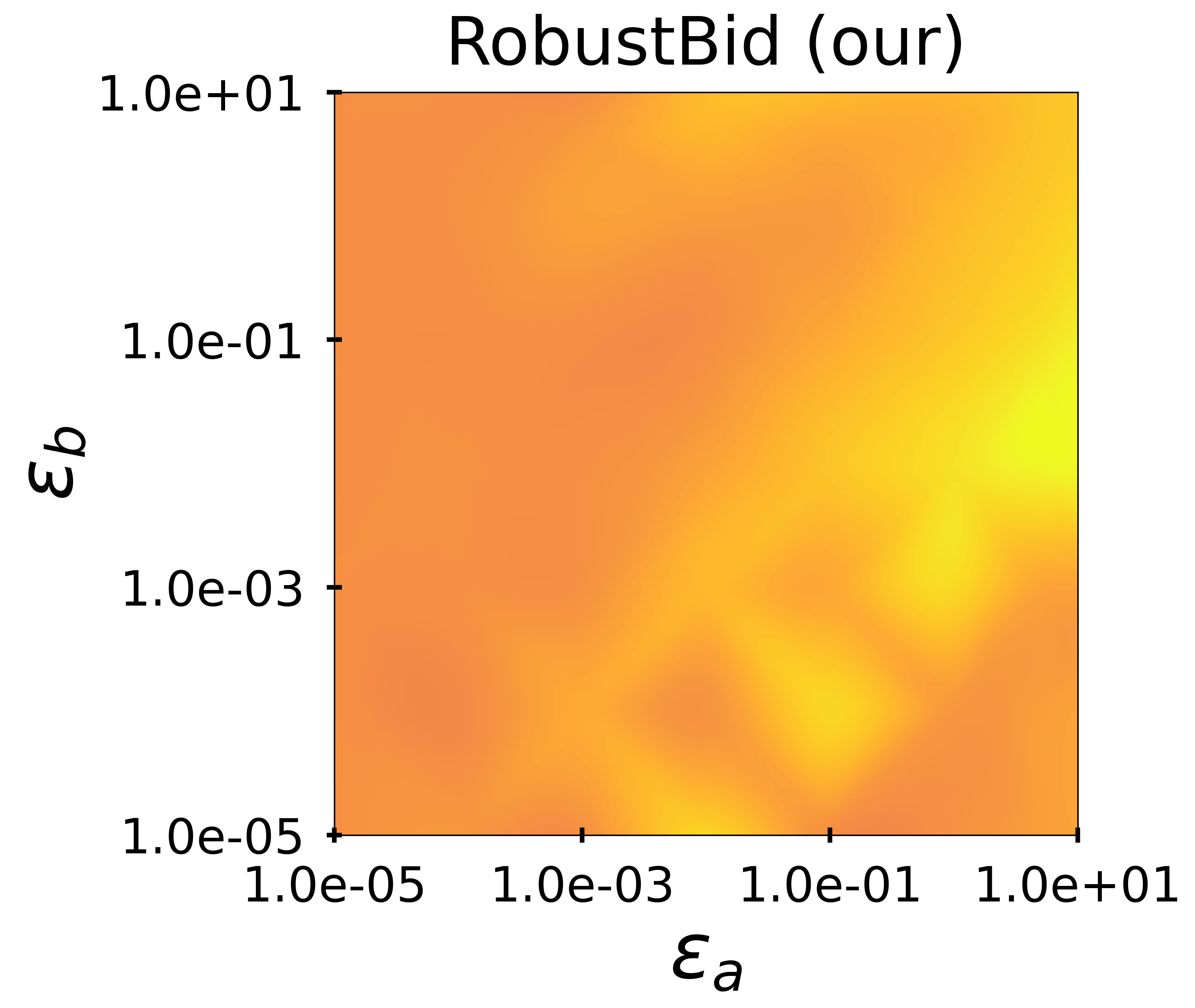}
    \end{minipage}
    \hfill
    \begin{minipage}[b]{0.15\linewidth}
        \centering
        \includegraphics[width=0.5\textwidth]{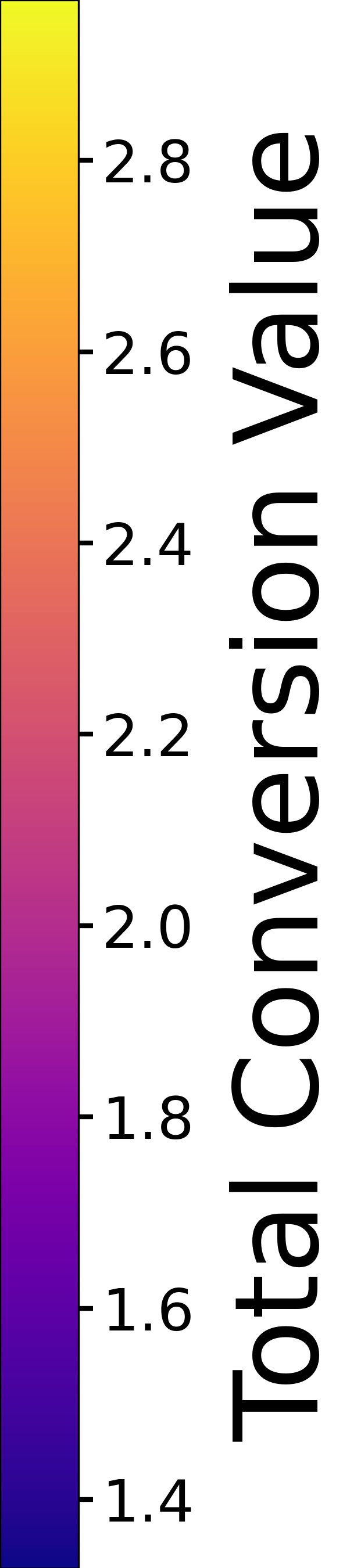}
    \end{minipage}
    
    
    \begin{minipage}[b]{0.33\linewidth}
        \centering
        \includegraphics[width=1.25\textwidth]{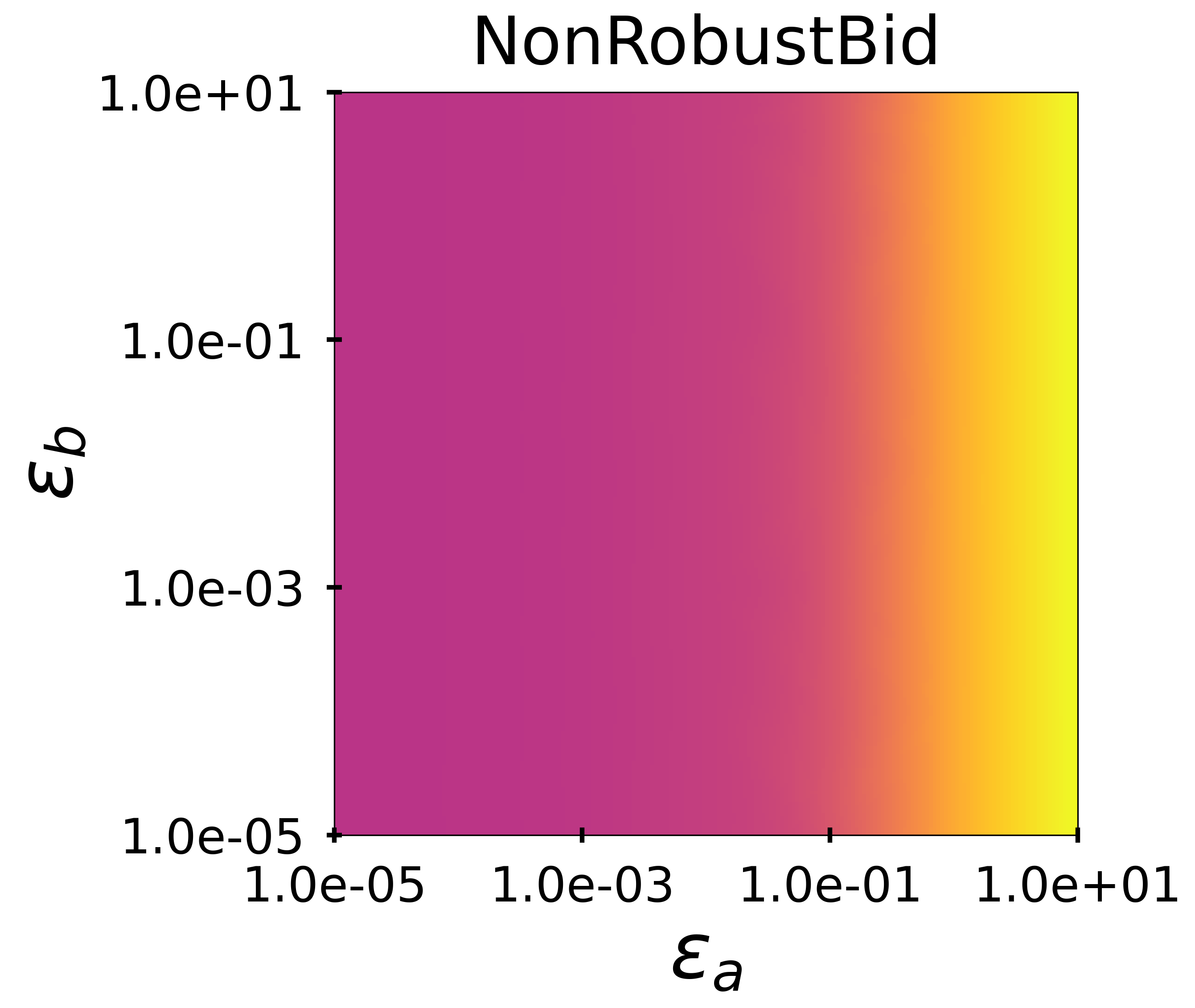}
    \end{minipage}
    \hfill
    \begin{minipage}[b]{0.33\linewidth}
        \centering
        \includegraphics[width=1.25\textwidth]{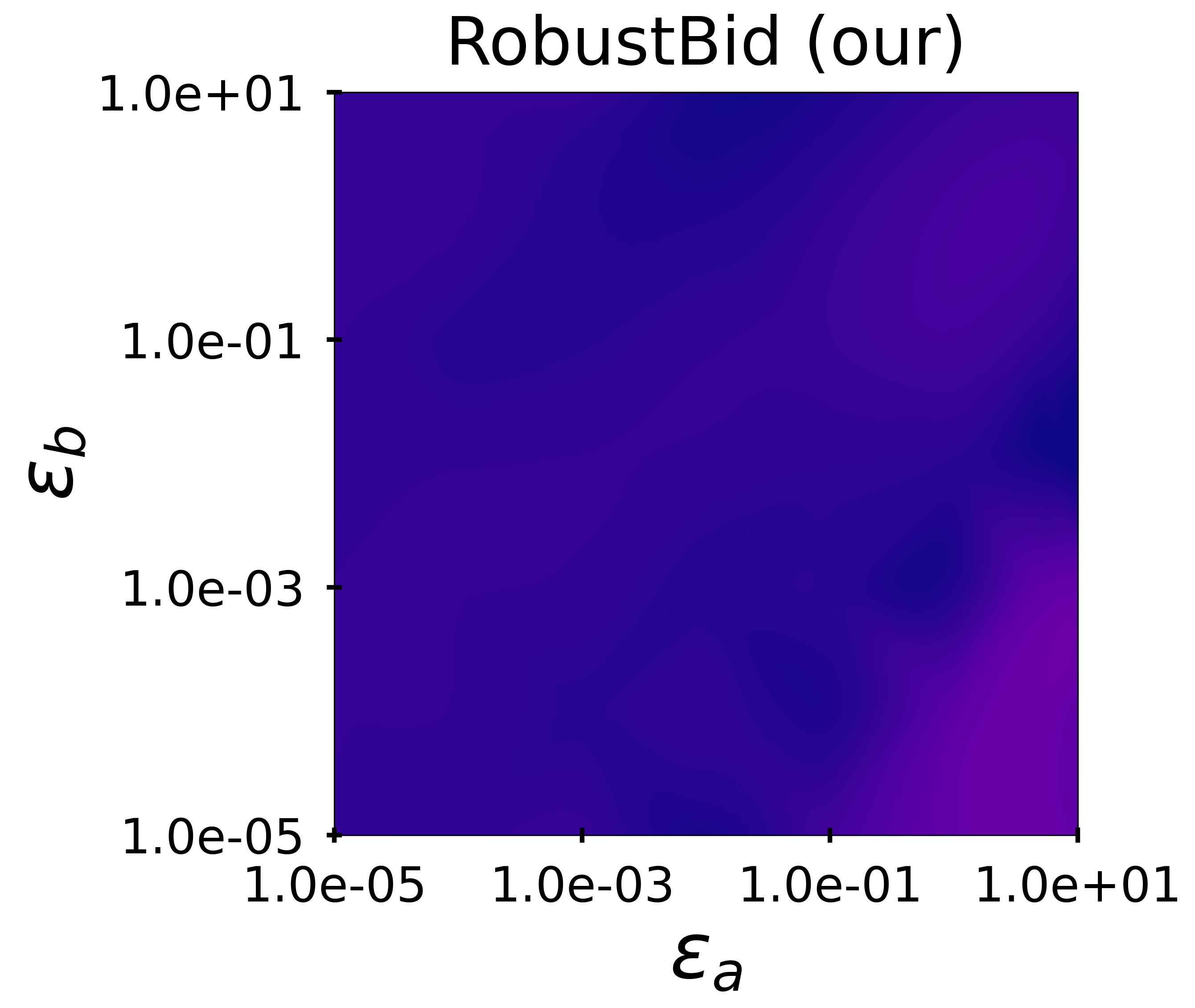}
    \end{minipage}
    \hfill
    \begin{minipage}[b]{0.15\linewidth}
        \centering
        \includegraphics[width=0.5\textwidth]{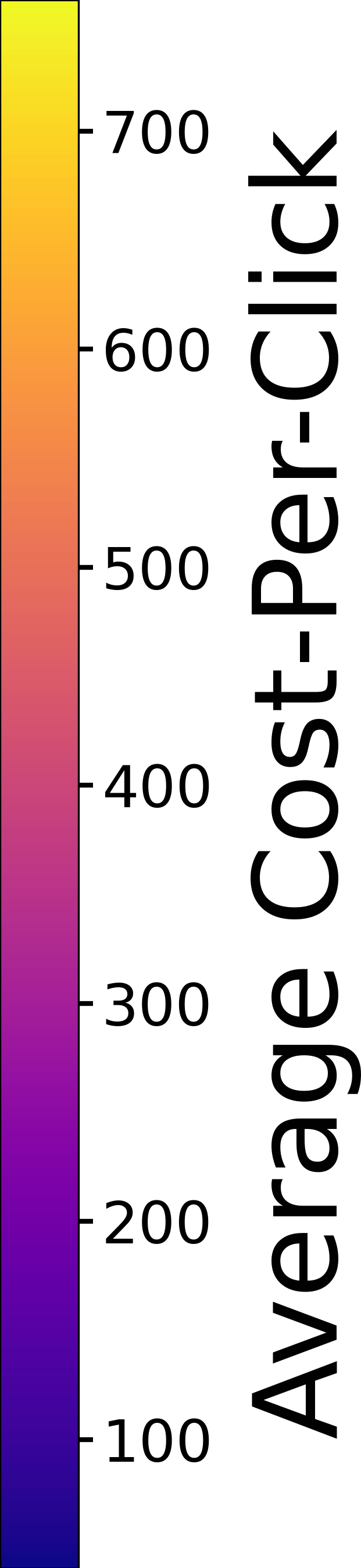}
    \end{minipage}
    
    \caption{Heatmaps with comparison $TCV$ and $CPC_{avg}$ for BAT dataset. 
    While $\varepsilon_a$ grows, \texttt{NonRobustBid} shows lower $TCV$ and higher $CPC_{avg}$. 
    At the same time, \texttt{RobustBid} metrics are nearly independent of $\varepsilon_a$ and $\varepsilon_b$.}
    \label{fig::CTRCVR_BAT}
\end{figure}

We analyzed the performance of our \texttt{RobustBid} approach compared to \texttt{NonRobustBid} across three datasets with varying $\varepsilon_a$ and $\varepsilon_b$, which refer to errors on $CTR$ and on $CVR$ respectively.
Comparing heatmap patterns across datasets, we observe that BAT (see Figure~\ref{fig::CTRCVR_BAT}) shows more stable behavior than iPinYou (see Figure~\ref{fig::CTRCVR_ipinyou}). 
This stability correlates with budget diversity: BAT has a wide budget range while iPinYou uses fixed budgets (see Table~\ref{table:parameters}).
On the synthetic dataset (see Figure~\ref{fig::CTRCVR_synth}), \texttt{RobustBid} consistently outperforms \texttt{NonRobustBid} when $\varepsilon_a$ and $\varepsilon_b$ are small enough.
The BAT dataset (Figure~\ref{fig::CTRCVR_BAT}) demonstrates strong results for \texttt{RobustBid}, maintaining stable metrics while consistently outperforming \texttt{NonRobustBid}.
This is especially significant as BAT most closely represents real-world advertising scenarios.
Across all three datasets and for the vast majority of $\varepsilon_a$ and $\varepsilon_b$  parameter pairs, \texttt{RobustBid} consistently achieves higher total conversions (TCV) and lower average cost-per-click ($CPC_{avg}$) than \texttt{NonRobustBid}, demonstrating the effectiveness of our robust approach.

\section{Limitations and future work}

Our approach works in offline setups, where a large amount of data is available, which can not be used in online setups.
The generalization of the proposed method to the online setup with controllers or RL frameworks is a promising research direction.
The incorporation of uncertainty sets based on cross-entropy loss in our method remains a challenging problem that can improve the interpretation of the uncertainty bounds.

\section{Conclusion}
This paper proposes the \texttt{RobustBid} algorithm that generates bids that are robust to uncertainties in CTR and CVR.
This algorithm is based on the robust modifications of the conversion maximization problem, incorporating uncertainty sets for individual CTR and CVR values and joint ones.  
The analytical solutions of the corresponding problems lead to the explicit formulas for bid, which naturally generalize the non-robust approach.
The extensive experiments on synthetic and industrial datasets confirm that \texttt{RobustBid} provides bids that are robust to perturbation in CTR/CVR.
The performance of the considered methods is evaluated through the total conversion rate and average cost per click.
According to these metrics, \texttt{RobustBid} gives a larger total conversion rate and a lower average cost per click than non-robust and robust competitors.

\bibliographystyle{unsrt} 
\bibliography{arxiv/main_arxiv}

\newpage
\appendix
\onecolumn
\section*{Appendix.} \label{sec:appendix}

\allowdisplaybreaks

\tableofcontents

\section{Solving internal optimization problem} \label{sec:internal_opt}

This section describes a solution for a helpful optimization problem for MSE loss: $\min_{\boldsymbol{a} \in \mathcal{U}^i}  \sum_{t=1}^T x_{t} \cdot a_{t} \cdot CVR_{t}$. As $\mathbf{m} = (x_{1} CVR_1, \ldots, x_T CVR_T)$ is a constant in this optimization problem and $\mathcal{U}^i  = \{ \mathbf{a} = (a_1, \ldots a_T), a_t \in [0,1] \quad \lvert \quad \  \frac{1}{2} \| \mathbf{a} - \bctr^i \|_2^2 \leq \varepsilon^i \}$ it is equal to solve next optimization problem: 
$$
\min _{ \frac{1}{2} \| \bctr - \bctr^i \|_2^2 \leq \varepsilon^i} \mathbf{m}^{T} \bctr.  
$$

To solve this optimization problem, the corresponding Lagrangian function is constructed:

$$
L(\bctr, \lambda) = \mathbf{x}^T \bctr+\lambda(\frac{1}{2} \| \bctr - \bctr^i \|_2^2 - \varepsilon^i).
$$

Let us examine the derivative of the Lagrangian function at the optimal point $(\bctr^{worst}, \lambda)$:

$$
\frac{\partial L}{\partial \bctr} (\bctr^{worst}, \lambda) = \mathbf{m} + \lambda (\bctr^{\text {worst }} - \bctr^i) = 0.
$$

From this equation, an expression for $\bctr^{\text{worst}}$ can be derived:

$$
\bctr^{\text {worst }} = \frac{\lambda \bctr^i - \mathbf{m}}{\lambda}.
$$

Subsequently, the constraint equation $\frac{1}{2}  \| \bctr^{\text{worst}} - \bctr^i\|^2_2 = \varepsilon$ is considered:

\begin{equation}
\frac{1}{2}  \| \frac{\lambda  \bctr^i - \mathbf{m}}{\lambda} - \bctr^i  \|^2_2 = 
\frac{1}{2} \| \frac{ \mathbf{m}}{\lambda}\|^2_2 = 
\varepsilon^i.
\end{equation}

Get and then use KKT conditions - $\lambda > 0$
\begin{equation}
\lambda = \frac{\|\mathbf{m}\|_2}{\sqrt{ 2 \varepsilon^i }},
\end{equation}

$$
\bctr_{\text {worst }}=\bctr^i-\frac{\mathbf{m}}{\|\mathbf{m}\|_{2}} \sqrt{ 2 \varepsilon^i}.
$$

This resulted in the robust counterpart of the initial inequality.

$$
b \leq \mathbf{m}^{T} \bctr^i - \|\mathbf{m}\|_{2} \sqrt{2 \varepsilon^i}.
$$

Notice that this optimization problem and its solution are used in both constraints of the original optimization problem.

\section{Robust problem reformulation and solution for CTR uncertainty} 
\label{sec:mse_theory}

This section contains a solution for the MSE loss optimization problem \ref{sec:mse_main}. The result of this section is the bid formula.

Consider the problem only for one item and use the next notation $\bctr^i = \bctr = (CTR_1, \ldots, CTR_T)$:
\begin{align*}
&\max _{x_{t}} \sum_{t=1}^T x_{t} \cdot CTR_{t} \cdot CVR_{t} \\
&\text { s.t. } \quad \sum_{t=1}^T x_{t} \cdot w p_{t} \leq B \\
&\quad \quad \quad \frac{\sum_{t=1}^T x_{t} \cdot w p_{t}}{\sum_{t=1}^T x_{t} \cdot C T R_{t}} \leq C \\
&\text { where } \quad 0 \leq x_{t} \leq 1, \quad \forall t \in [1, T].
\end{align*}

Robust formulation means maximizing the expected number of clicks in a worthwhile situation (here and later in the paper, the notation of $i$ is never used):
$$\max_{x_t} \quad \min_{\mathcal{U}^i} \quad \sum_{t=1}^T x_t CTR_t CVR_t = \max_{x_t} \quad \sum_{t=1}^T x_t CTR_{t, \text{worst}} CVR_t,$$
which can be rewritten as
\begin{align*}
&\max _{\mathbf{x}, \mathbf{y}, z} ~~z \\
&\text { s.t. }\quad  y_{t}=x_{t} \cdot C V R_{t} \quad \forall t \\
&\hspace{11mm} z \leq \mathbf{y}^{T} \bctr - \alpha\|\mathbf{y}\|_{2} \\
&\hspace{10mm}\sum_{t=1}^T x_{t} \cdot w p_{t} \leq B \\
&\hspace{10mm}\sum_{t=1}^T x_{t} \cdot w p_{t} \leq C \cdot [\mathbf{x}^{T} \bctr - \alpha\|\mathbf{x}\|_{2}  ] \\
&\hspace{11mm} 0 \leq x_{t} \leq 1, \forall t\\
&\text{where } \alpha=\sqrt{2 \varepsilon^i}
\end{align*}

\subsection*{Dual Problem}

The original problem from~\cite{pid2019} was solved by the dual method. 
We make up a dual too, and the formula for the optimal bid is found. 
In the main part, we use the notation of $p,q$ for dual variables as it is commonly used in literature. However, in this appendix, we exploit mapping from $p, q$ to $\gamma, u_0$.
Firstly, we have to rewrite some constraints in terms of conic terms. Using fact that ${\alpha}||\mathbf{y}||_2 \leq -z + \mathbf{y}^T \bctr$ is equal to
$\begin{pmatrix}
    \alpha \mathbf{ y} \\ -z +  \mathbf{y}^T \bctr
\end{pmatrix} \in K_2 = \{ (\bvectorx, c) \quad | \quad  ||\bvectorx||_2 \leq  c\}$.
Consider the Lagrange function for our problem:
\begin{eqnarray*}
    L(\mathbf{x}, \mathbf{y}, z, \delta, (\bvectormu, \mu_0)^T, \gamma, (\mathbf{u}, -u_0)^T, \boldsymbol{\nu_1}, \boldsymbol{\nu_2}) &=& 
    z + \sum_{t=1}^{T} \delta_t (y_t - x_t CVR_t) \\ && - \gamma (\sum_{t=1}^{T} x_t wp_t - B) + 
    (\bvectormu,  \mu_0)
    \begin{pmatrix}
        \alpha \mathbf{y} \\ -z +  \mathbf{y}^T \bctr
    \end{pmatrix} 
     \\ &&+ (\mathbf{u}, u_0)
     \begin{pmatrix}
        C\alpha\mathbf{x} \\ -\sum_{t=1}^T x_t wp_t + C \bvectorx^T \bctr
    \end{pmatrix} 
    + \bvectorx^T \boldsymbol{\xi}_1 - \boldsymbol{\xi}_2^T (\bvectorx - \overline{1}).
\end{eqnarray*}
Due to optimality, we have constraints,
$$
    \begin{pmatrix}
        \bvectormu \\ \mu_0
    \end{pmatrix} \in K_2^* = K_2,
    \begin{pmatrix}
        \mathbf{u} \\ u_0
    \end{pmatrix} \in K_2, \gamma \geq 0, \boldsymbol{\xi}_1 \geq 0, \boldsymbol{\xi}_2 \geq 0.
$$

Compute gradients for optimality:
\begin{eqnarray*}
    \nabla_{x_t} L &=& - \delta_t CVR_t - \gamma wp_t + C \alpha u_t -  wp_t u_0 + C u_0 \text{CTR}_t + (\xi_1)_t -(\xi_2)_t \\
    \nabla_{y_t} L &=& \delta_t + \alpha \mu_t + \mu_0 \text{CTR}_t \\
    \nabla_z L &=& 1 - \mu_0.
\end{eqnarray*}

From $\nabla_{x_t} L = 0, \nabla_{y_t} L = 0, \nabla_z L = 0$ we get $\mu_0 = 1, \delta_t +  \alpha \mu_t = - \text{CTR}_t$. Then,
\begin{equation}
\label{robust_dual_stright}
\begin{aligned}
    & \min _{\boldsymbol{\delta} \in R^T, \boldsymbol{\mu} \in R^T, \mathbf{u} \in R^T, \boldsymbol{\xi}_1 \in R^T, \boldsymbol{\xi}_2 \in R^T, \gamma \in R, u_0 \in R} \gamma B+\sum_{t=1}^T (\xi_2)_t \\
    & \delta_t + \alpha \mu_t = - \text{CTR}_t \\
    & \delta_t {CVR}_t + \gamma \cdot w p_t +  C \alpha u_t + u_0 \cdot (-C \cdot \text{CTR}_t+wp_t) - (\xi_1)_t + (\xi_2)_t=0 \\
    & \|\bvectormu\|_2 \leq 1, \quad \|\mathbf{u}\|_2 \leq u_0 \\
    & \boldsymbol{\xi}_1 \geq 0, \quad \boldsymbol{\xi}_2 \geq 0, \quad \gamma \geq 0
\end{aligned}
\end{equation}

We can simplify the problem by using an idea, which we can get from considering cases of $({\xi}_1)_t, ({\xi}_2)_t$.

Consider $\nabla_{x_t} L = 0$, which is equal to $- \delta_t CVR_t - \gamma wp_t +  C\alpha u_t - wp_t u_0 + C u_0 \text{CTR}_t + (\xi_1)_t = (\xi_2)_t$.
So if $(\xi_2)_t > 0$ then by complementary slackness $x_t = 1, (\xi_1)_t = 0$. If $(\xi_2)_t = 0$, then $x_t=0$. The set where $(\xi_2)_t < 0$ is not considered, because it is not feasible due to the conditions. 
In this way,
$(\xi_2)_t = max(0, - \delta_t CVR_t - \gamma wp_t +  \alpha u_t - wp_t u_0 + C u_0 \text{CTR}_t)$.

\begin{equation}
\begin{gathered}
\label{robust_formulation_exp}
\min _{\boldsymbol{\delta} \in R^T, \bvectormu \in R^T, \mathbf{u} \in R^T, \gamma \in R, u_0 \in R}  ( \gamma B+  \sum_{t=1}^T max(0, -\delta_t CVR_t - \gamma wp_t - C \alpha  u_t - u_0 wp_t + C u_0 \text{CTR}_t)  ),\\
 \|\boldsymbol{\delta} + \bctr\|_2 \leq \alpha ,  \quad  \|\mathbf{u}\|_2 \leq u_0, \gamma \geq 0.
\end{gathered}
\end{equation}

\subsection*{Robust problem solution}
We denote active indexes (indexes when function $\max(0, ...) > 0$) as $\mathcal{T}$ for simplicity.

\subsection{How to find $\mathbf{\delta}$}
Solve problem on $\delta$:

$$
\min _{\boldsymbol{\delta} \in R^T, \bvectormu \in R^T, \mathbf{u} \in R^T,  \gamma \in R, u_0 \in R}  ( \gamma B+  \sum\limits_{t\in \mathcal{T}}(-\delta_t CVR_t - \gamma wp_t -  C \alpha u_t - u_0 wp_t +C u_0 \text{CTR}_t)  ).
$$

Define helpful $\tilde{\mathbf{c}} = (\tilde{c_1}, ..., \tilde{c_T})$, the solution is on bound $\alpha \|\boldsymbol{\delta} + \bctr\|_2 \leq \alpha$. Thus,
\begin{equation*}
  \tilde{c}_t=\begin{cases}
    CVR_t, & t\in\mathcal{T},\\
    0, & \text{otherwise}.
  \end{cases}
\end{equation*}

Then we have an intermediate problem only with the summand depending on $\boldsymbol{\delta}$:

$$
\begin{cases}
    \min _{\boldsymbol{\delta} \in R^T, \bvectormu \in R^T, \mathbf{u} \in R^T,  \gamma \in R, u_0 \in R} \sum\limits_{t\in \mathcal{T}}-\delta_t CVR_t  + P, \\
     \|\boldsymbol{\delta} + \bctr\|_2 \leq \alpha.
\end{cases}
$$

Where $P$ is independent with $\boldsymbol{\delta}$ and have Lagrangian: $L = - \boldsymbol{\delta}^T \tilde{\mathbf{c}} + \lambda ( ||\boldsymbol{\delta} + \bctr||_2 - \alpha)$. In this way,

$$\frac{\partial L}{\partial \boldsymbol{\delta}} = -\tilde{\mathbf{c}} + \lambda \frac{\boldsymbol{\delta} + \bctr}{\| \boldsymbol{\delta} + \bctr \|_2} = 0.$$

$$\lambda   \frac{\boldsymbol{\delta} + \bctr}{\| \boldsymbol{\delta} + \bctr \|_2} =\tilde{\mathbf{c}}.$$ 

$$\lambda  = \| \tilde{\mathbf{c}}\|_2.$$

$$\boldsymbol{\delta} + \bctr = \frac{\tilde{\mathbf{c}}}{\| \tilde{\mathbf{c}}\|_2} \| \boldsymbol{\delta} + \bctr \|_2 .$$

$$\boldsymbol{\delta} = - \bctr + \frac{\tilde{\mathbf{c}}}{\| \tilde{\mathbf{c}}\|_2} \| \boldsymbol{\delta} + \bctr \|_2.$$







The solution is on bound, because we minimia ze linear function on a sphere, then  $ \|\boldsymbol{\delta}+\bctr \|_2=\alpha$ and obtain:
$ \boldsymbol{\delta} = - \bctr + \frac{\alpha }{ \| \tilde{\mathbf{c}}\|_2} \tilde{\mathbf{c}}$.

Optimal solution for $\boldsymbol{\delta}$ is:

\begin{align*}
 \delta_t^*=\begin{cases}
    - \text{CTR}_t + \frac{\alpha CVR_t}{ \sqrt{\sum_{l \in \mathcal{T}} CVR^2_l}}, & t\in\mathcal{T};\\
    - \text{CTR}_t, & \text{otherwise}.
  \end{cases}
\end{align*}

\subsection{$u_t$}

\begin{align*}
    \min _{\boldsymbol{\delta} \in R^T, \bvectormu \in R^T, \mathbf{u} \in R^T,  \gamma \in R, u_0 \in R}  &\left( \gamma B+  \sum_{i=1}^T \max(0, - {\delta}_t CVR_t - \gamma wp_t -  C \alpha u_t - u_0 wp_t +C u_0 \text{CTR}_t)  \right), \\
 &\|\boldsymbol{\delta} + \bctr \|_2 \leq \alpha, \quad \|\mathbf{u}\|_2 \leq u_0 , \quad \gamma \geq 0.
\end{align*}

Define helpful variable $A_t = - C \alpha u_t - u_0 wp_t +C u_0  {CTR}$. Note that $A_t > 0$. To proof it consider the optimal pair $(\mathbf{u}^*, u_0)$ and a scalar $\beta > 1$. Then the pair $\beta(u^*, u_0)$ is feasible and more suitable, as it results in a smaller total sum. 

To simplify our notation, let us define: $d = - wp_t +\text{CTR}_t C $.

Let us consider the Lagrangian function for each auction $t$:

$$\mathcal{L}_t(u) = - C \alpha u_t+du_0 + \lambda ( \un - u_0).$$

Differentiating with respect to $u_t$, we obtain:

$$\frac{\partial\mathcal{L}}{\partial u_t} = - C \alpha + \lambda  \frac{u_t}{\un} = 0. $$

This implies that $u_t = \frac {C \alpha}{\lambda}\un$ for $t\in\mathcal{T}$. We shall now treat $u_0$ as a fixed hyperparameter. It is noteworthy that all positive $u_t$ for $t\in\mathcal{T}$ are equal. Our objective is to demonstrate that $\un = u_0$, indicating that the solution lies on the boundary.

To explain why $u_t = 0$ for $t\notin \mathcal{T}$, let us rewrite the condition.
If there exists a $j^{*} \notin \mathcal{T}$ such that $u_j \neq 0$, then $u_{j}$ can be set to 0, and for some $\hat{t}$, this value can be redistributed, resulting in a new value $\hat{u}_{\hat{t}} = \sqrt{u_{\hat{t}}^2 + u_{j^*}^2}$. This redistribution maintains the same $\un$.
To address why $ \un =  \sqrt{\sum_{t \in \mathcal{T}} u_t^2} = u_0$ for $t \in \mathcal{T}$: With $u_0$ fixed, if $u_t$ could be larger, the corresponding $A_t$ would be smaller, giving a better solution to the minimization problem. This reasoning leads us to conclude that the optimal solution must lie on the boundary.

Utilizing the definition of the Euclidean norm $||.||_2$, we can derive:
$\lambda = C\alpha\sqrt{\mid \mathcal{T}\mid}$, $u_t = \un \sqrt{\mid \mathcal{T}\mid}^{-1}$, where $|\mathcal{T}|$ denotes the cardinality of set $\mathcal{T}$.
Now, we can formulate the complete expression for $u_t$:

\begin{align*}
 u_t=\begin{cases}
    \frac{1}{\sqrt{\mid\mathcal{T}\mid}}u_0, &  t \in \mathcal{T};  \\
    0, & t\notin \mathcal{T}.
  \end{cases}
\end{align*}

\subsection{$wp$}
Let us now examine the third condition from equation \ref{robust_dual_stright}, and recall that the dual variables $-\boldsymbol{\xi}_1$ and $-\boldsymbol{\xi}_2$ correspond to the constraints $x_t \geq 0$ and $x_t \leq 1$, respectively:
$$\delta_t C V R_t+\gamma \cdot w p_t + C\alpha u_t-u_0 (C \cdot \text{CTR}_t - wp_t) - (\xi_1)_t + (\xi_2)_t = 0.$$

We analyze different cases of $x_t$ by considering the complementary slackness conditions: $\bvectorx^T \boldsymbol{\xi_1} = 0$ and $\boldsymbol{\xi}_2(\bvectorx - \overline{1}) = 0$.
Additionally, we utilize the lower bounds $\boldsymbol{\xi}_1 \geq 0, \boldsymbol{\xi}_2 \geq 0$ from equation \ref{robust_dual_stright}:

\begin{enumerate}
    \item $x_t = 1 \Longrightarrow (\xi_2)_t \geq 0, (\xi_1)_t=0$;
    \item  $x_t = 0 \Longrightarrow (\xi_2)_t = 0, (\xi_1)_t \geq 0$.
\end{enumerate}

$$\delta_t CVR_t + C\alpha u_t - u_0 C \; {CTR}_t  + wp_t (\gamma + u_0) = (\xi_1)_t - (\xi_2)_t. $$

1. $x_t=1$ then $(\xi_1)_t - (\xi_2)_t \leq 0$ obtain $\delta_tCVR_t + \alpha u_t  -u_0 C\cdot \text{CTR}_t  \leq -wp_t(\gamma + u_0)$

Use $\gamma \geq 0, \alpha > 0$ and $u_0 > 0$ to satisfy condition $ \un \leq u_0$. Then get 
$$ \frac{-\delta_t CVR_t-C\alpha u_t + C u_0 \cdot \text{CTR}_t }{\gamma + u_0}\geq wp_t.$$

2. $x_t = 0$ then $(\xi_1)_t - (\xi_2)_t \geq 0$ and then:
$$ \frac{- \delta_t CVR_t-C\alpha u_t +C u_0 \cdot \text{CTR}_t }{\gamma +u_0}\leq wp_t.$$



So, the conclusion is a formula for bid as in \cite{pid2019}:
\begin{multline*}   
bid_t = \frac{CTR_t CVR_t}{\gamma +u_0} +\frac {u_0 C \cdot CTR_t} {\gamma+u_0}+
\begin{cases}
     - \frac {\alpha}{\gamma+u_0} (\frac{Cu_0}{\sqrt{\mid\mathcal{T}\mid}} + \frac{ CVR_t^2}{ \sqrt{\sum_{l \in \mathcal{T}} CVR^2_l}} ) \quad, t \in \mathcal{T};\\
     0,
    \quad t \notin \mathcal{T}.
\end{cases}
\end{multline*}

\subsection{Active indices $\mathcal{T}$}

Inactive indices are those for which the maximum opened at 0. 
$$0 \geq -\delta_t^* CVR_t - \gamma wp_t - C\alpha u_t^* -  u_0 wp_t+ u_0 C \cdot \text{CTR}_t.$$
In this case, $\delta_t^*= -\text{CTR}_t$, $u_t^*=0$. In the case of inactive index, $CTR_t CVR_t+C u_0 CTR_t=(\gamma + u_0)bid_t$, and the equation above is true when $bid_t \leq wp_t$ and the auction is lost.
Active indices are those for which maximum opened not at 0.
$$-\delta_t^* CVR_t - \gamma wp_t - C\alpha u_t^* -  u_0 wp_t+ u_0 C \cdot \text{CTR}_t \geq 0.$$

$$ \text{CTR}_t CVR_t - \frac{\alpha CVR_t}{ \sqrt{\sum_{l \in \mathcal{T}} CVR^2_l}} CVR_t - \gamma wp_t -C \alpha \frac{1}{\sqrt{\mid\mathcal{T}\mid}}u_0 -  u_0 wp_t+ u_0 C \cdot \text{CTR}_t \geq 0.$$

$$ \text{CTR}_t CVR_t + u_0 C \cdot \text{CTR}_t - \alpha( \frac{ CVR_t^2}{ \sqrt{\sum_{l \in \mathcal{T}} CVR^2_l}} + \frac{1}{\sqrt{\mid\mathcal{T}\mid}}Cu_0)\geq  \gamma wp_t +  u_0 wp_t .$$


In this case $\text{CTR}_t CVR_t + u_0 C \cdot \text{CTR}_t - \alpha( \frac{ CVR_t^2}{ \sqrt{\sum_{l \in \mathcal{T}} CVR^2_l}} + \frac{1}{\sqrt{\mid\mathcal{T}\mid}}Cu_0) = (\gamma + u_0) bid_t$, and the equation above is true when $bid_t \geq wp_t$ and the auction is won.

Overall, the next lemma can be formulated.

\begin{lemma}[The bid for the CTR uncertainty in MSE terms]
\label{app:lemma_mse_ctr}
    For the robust optimization problem \ref{opt:robust_optimization} the closed formula for the bid can be obtained \ref{eq::robust_bid_ctr} with active indexes defined as \ref{inactive_index_formula}
\end{lemma}

\section{Robust problem reformulation and solution for CVR uncertainty}
\label{sec::appendix_only_cvr}

This section contains solution for the MSE loss optimization problem \ref{sec:mse_main} as Appendix \ref{sec:mse_theory}, however this section solves problem for only CVR uncertainty as it was mentioned in the main part \ref{sec::main_robustbid}. 
Most steps are repeated or simplified from Appendix \ref{sec:mse_theory}, although we focus on the distinction between these solutions. The result of this section is the bid formula too. 

Consider the problem only for one item and use the next notation $\bctr^i = \bctr = (CTR_1, \ldots, CTR_T)$:
\begin{align*}
&\max _{x_{t}} \sum_{t=1}^T x_{t} \cdot CTR_{t} \cdot CVR_{t} \\
&\text { s.t. } \quad \sum_{t=1}^T x_{t} \cdot w p_{t} \leq B \\
&\quad \quad \quad \frac{\sum_{t=1}^T x_{t} \cdot w p_{t}}{\sum_{t=1}^T x_{t} \cdot C T R_{t}} \leq C \\
&\text { where } \quad 0 \leq x_{t} \leq 1, \quad \forall t \in [1, T].
\end{align*}

Robust formulation means maximizing the expected number of clicks in a worthwhile situation (here and later in the paper, the notation of $i$ is never used):
$$\max_{x_t} \quad \min_{\mathcal{U}^i} \quad \sum_{t=1}^T x_t CTR_t CVR_t = \max_{x_t} \quad \sum_{t=1}^T x_t CTR_t CVR_{t, \text{worst}},$$
which can be rewritten as
\begin{align*}
&\max _{\mathbf{x}, \mathbf{y}, z} ~~z \\
&\text { s.t. }\quad  y_{t}=x_{t} \cdot CTR_{t} \quad \forall t \\
&\hspace{11mm} z \leq \mathbf{y}^{T} \mathbf{CVR} - \alpha\|\mathbf{y}\|_{2} \\
&\hspace{10mm}\sum_{t=1}^T x_{t} \cdot w p_{t} \leq B \\
&\hspace{10mm}\sum_{t=1}^T x_{t} \cdot (w p_{t} - C \cdot CTR_t) \leq 0  \\
&\hspace{11mm} 0 \leq x_{t} \leq 1, \forall t\\
&\text{where } \alpha=\sqrt{2 \varepsilon^i}
\end{align*}

\subsection*{Dual Problem}

This section exploits the idea of conic constraints too and formulates a dual problem, which is necessary for determining the bid function.
However, in this appendix, we exploit mapping from $p, q$ to $\gamma, u_0$.

Considerthe  Lagrange function for our problem:
\begin{eqnarray*}
    L(\mathbf{x}, \mathbf{y}, z, \delta, (\bvectormu, \mu_0)^T, \gamma, (\mathbf{u}, -u_0)^T, \boldsymbol{\nu_1}, \boldsymbol{\nu_2}) &=& 
    z + \sum_{t=1}^{T} \delta_t (y_t - x_t CVR_t)  - \gamma (\sum_{t=1}^{T} x_t wp_t - B) + 
    (\bvectormu,  \mu_0)
    \begin{pmatrix}
        \alpha \mathbf{y} \\ -z +  \mathbf{y}^T \bctr
    \end{pmatrix} 
     \\ &&-  u_0 \sum_{t=1}^T x_t (wp_t - C \cdot CTR_t)
    + \bvectorx^T \boldsymbol{\xi}_1 - \boldsymbol{\xi}_2^T (\bvectorx - \overline{1}).
\end{eqnarray*}
Due to optimality, we have constraints,
$$
    \begin{pmatrix}
        \bvectormu \\ \mu_0
    \end{pmatrix} \in K_2^* = K_2,
     \gamma, u_0 \geq 0, \boldsymbol{\xi}_1 \geq 0, \boldsymbol{\xi}_2 \geq 0.
$$

Compute gradients for optimality:
\begin{eqnarray*}
    \nabla_{x_t} L &=& - \delta_t CTR_t - \gamma wp_t -  u_0(wp_t - C \cdot CTR_t) + (\xi_1)_t -(\xi_2)_t \\
    \nabla_{y_t} L &=& \delta_t + \alpha \mu_t + \mu_0 \text{CVR}_t \\
    \nabla_z L &=& 1 - \mu_0.
\end{eqnarray*}

From $\nabla_{x_t} L = 0, \nabla_{y_t} L = 0, \nabla_z L = 0$ we get $\mu_0 = 1, \delta_t +  \alpha \mu_t = - \text{CVR}_t$. 
Then,

\begin{equation}
\label{robust_dual_stright_cvr}
\begin{aligned}
    & \min _{\boldsymbol{\delta} \in R^T, \boldsymbol{\mu} \in R^T, \mathbf{u} \in R^T, \boldsymbol{\xi}_1 \in R^T, \boldsymbol{\xi}_2 \in R^T, \gamma \in R, u_0 \in R} \gamma B+\sum_{t=1}^T (\xi_2)_t \\
    & \delta_t + \alpha \mu_t = - \text{CVR}_t \\
    & \delta_t {CTR}_t + \gamma \cdot w p_t +  u_0 \cdot (-C \cdot \text{CTR}_t+wp_t) - (\xi_1)_t + (\xi_2)_t=0 \\
    & \|\bvectormu\|_2 \leq 1,  \quad \boldsymbol{\xi}_1 \geq 0, \quad \boldsymbol{\xi}_2 \geq 0, \quad \gamma \geq 0, \quad u_0 \geq 0
\end{aligned}
\end{equation}

We can simplify the problem by using an idea, which we can get from considering cases of $({\xi}_1)_t, ({\xi}_2)_t$.

Consider $\nabla_{x_t} L = 0$, which is equal to $- \delta_t CTR_t - \gamma wp_t  - wp_t u_0 + C u_0 \text{CTR}_t + (\xi_1)_t = (\xi_2)_t$.
So if $(\xi_2)_t > 0$ then by complementary slackness $x_t = 1, (\xi_1)_t = 0$. If $(\xi_2)_t = 0$, then $x_t=0$. The set where $(\xi_2)_t < 0$ is not considered, because it is not feasible due to the conditions. 
In this way,
$(\xi_2)_t = max(0, - \delta_t CTR_t - \gamma wp_t - wp_t u_0 + C u_0 \text{CTR}_t)$.

\begin{equation}
\begin{gathered}
\label{robust_formulation_exp_cvr}
\min _{\boldsymbol{\delta} \in R^T, \bvectormu \in R^T, \mathbf{u} \in R^T, \gamma \in R, u_0 \in R}  ( \gamma B+  \sum_{t=1}^T max(0, -\delta_t CVR_t - \gamma wp_t  - u_0 wp_t + C u_0 \text{CTR}_t)  ),\\
 \|\boldsymbol{\delta} + \mathbf{CVR}\|_2 \leq \alpha ,  \quad u_0 \geq 0, \quad \gamma \geq 0.
\end{gathered}
\end{equation}

\subsection*{MSE robust problem solution}
We denote active indexes (indexes when function $max(0, ...) > 0$) as $\mathcal{T}$ for simplicity.

\subsection{How to find $\mathbf{\delta}$}
Solve problem on $\delta$:

$$
\min _{\boldsymbol{\delta} \in R^T, \bvectormu \in R^T, \mathbf{u} \in R^T,  \gamma \in R, u_0 \in R}  ( \gamma B+  \sum\limits_{t\in \mathcal{T}}(-\delta_t CTR_t - \gamma wp_t - u_0 wp_t +C u_0 \text{CTR}_t)  ).
$$

Define helpful $\tilde{\mathbf{c}} = (\tilde{c_1}, ..., \tilde{c_T})$, the solution is on bound $\alpha \|\boldsymbol{\delta} + \mathbf{CVR}\|_2 \leq \alpha$. Thus,
\begin{equation*}
  \tilde{c}_t=\begin{cases}
    CTR_t, & t\in\mathcal{T},\\
    0, & \text{otherwise}.
  \end{cases}
\end{equation*}

Then we have an intermediate problem only with the summand depending on $\boldsymbol{\delta}$:

$$
\begin{cases}
    \min _{\boldsymbol{\delta} \in R^T, \bvectormu \in R^T, \mathbf{u} \in R^T,  \gamma \in R, u_0 \in R} \sum\limits_{t\in \mathcal{T}}-\delta_t CTR_t  + P, \\
     \|\boldsymbol{\delta} + \mathbf{CVR}\|_2 \leq \alpha.
\end{cases}
$$

Where $P$ is independent with $\boldsymbol{\delta}$ and have Lagrangian: $L = - \boldsymbol{\delta}^T \tilde{\mathbf{c}} + \lambda ( ||\boldsymbol{\delta} + \mathbf{CVR}||_2 - \alpha)$. In this way,

$$\frac{\partial L}{\partial \boldsymbol{\delta}} = -\tilde{\mathbf{c}} + \lambda \frac{\boldsymbol{\delta} + \mathbf{CVR}}{\| \boldsymbol{\delta} + \mathbf{CVR} \|_2} = 0.$$

$$\lambda   \frac{\boldsymbol{\delta} + \mathbf{CVR}}{\| \boldsymbol{\delta} + \mathbf{CVR} \|_2} =\tilde{\mathbf{c}}.$$ 

$$\lambda  = \| \tilde{\mathbf{c}}\|_2.$$

$$\boldsymbol{\delta} + \mathbf{CVR} = \frac{\tilde{\mathbf{c}}}{\| \tilde{\mathbf{c}}\|_2} \| \boldsymbol{\delta} + \mathbf{CVR} \|_2 .$$

$$\boldsymbol{\delta} = - \mathbf{CVR} + \frac{\tilde{\mathbf{c}}}{\| \tilde{\mathbf{c}}\|_2} \| \boldsymbol{\delta} + \mathbf{CVR} \|_2.$$

The solution is on bound, because we minimize a linear function on a sphere, then  $ \|\boldsymbol{\delta}+ \mathbf{CVR} \|_2=\alpha$ and obtain:
$ \boldsymbol{\delta} = - \mathbf{CVR} + \frac{\alpha }{ \| \tilde{\mathbf{c}}\|_2} \tilde{\mathbf{c}}$.

Optimal solution for $\boldsymbol{\delta}$ is:

\begin{align*}
 \delta_t^*=\begin{cases}
    - \text{CVR}_t + \frac{\alpha CTR_t}{ \sqrt{\sum_{l \in \mathcal{T}} CTR^2_l}}, & t\in\mathcal{T};\\
    - \text{CVR}_t, & \text{otherwise}.
  \end{cases}
\end{align*}

\subsection{$wp$}
Let us now examine the third condition from equation \ref{robust_dual_stright_cvr}, and recall that the dual variables $-\boldsymbol{\xi}_1$ and $-\boldsymbol{\xi}_2$ correspond to the constraints $x_t \geq 0$ and $x_t \leq 1$, respectively:
$$\delta_t C T R_t+\gamma \cdot w p_t - u_0 (C \cdot \text{CTR}_t - wp_t) - (\xi_1)_t + (\xi_2)_t = 0.$$

We analyze different cases of $x_t$ by considering the complementary slackness conditions: $\bvectorx^T \boldsymbol{\xi_1} = 0$ and $\boldsymbol{\xi}_2(\bvectorx - \overline{1}) = 0$.
Additionally, we utilize the lower bounds $\boldsymbol{\xi}_1 \geq 0, \boldsymbol{\xi}_2 \geq 0$ from equation \ref{robust_dual_stright_cvr}:

\begin{enumerate}
    \item $x_t = 1 \Longrightarrow (\xi_2)_t \geq 0, (\xi_1)_t=0$;
    \item  $x_t = 0 \Longrightarrow (\xi_2)_t = 0, (\xi_1)_t \geq 0$.
\end{enumerate}

$$\delta_t CTR_t + - u_0 C \; {CTR}_t  + wp_t (\gamma + u_0) = (\xi_1)_t - (\xi_2)_t. $$

1. $x_t=1$ then $(\xi_1)_t - (\xi_2)_t \leq 0$ obtain $\delta_t CTR_t  -u_0 C\cdot \text{CTR}_t  \leq -wp_t(\gamma + u_0)$

Use $\gamma \geq 0, \alpha > 0$ and $u_0 > 0$. Then get 
$$ \frac{-\delta_t CTR_t + C u_0 \cdot \text{CTR}_t }{\gamma + u_0}\geq wp_t.$$

2. $x_t = 0$ then $(\xi_1)_t - (\xi_2)_t \geq 0$ and then:
$$ \frac{- \delta_t CTR_t-\alpha u_t +C u_0 \cdot \text{CTR}_t }{\gamma +u_0}\leq wp_t.$$

So, the conclusion is a formula for bid as in \cite{pid2019}:
\begin{equation}
\label{app::bid_cvr}
bid_t = \frac{CTR_t CVR_t}{\gamma +u_0} +\frac {u_0 C \cdot CTR_t} {\gamma+u_0}+
\begin{cases}
     - \frac {\alpha}{\gamma+u_0} \frac{ CTR_t}{ \sqrt{\sum_{l \in \mathcal{T}} CTR^2_l}}  \quad, t \in \mathcal{T};\\
     0,
    \quad t \notin \mathcal{T}.
\end{cases}
\end{equation}

\subsection{Active indices $\mathcal{T}$}
Inactive indices are those for which the maximum opened at 0. 
$$0 \geq -\delta_t^* CVTR_t - \gamma wp_t  -  u_0 wp_t+ u_0 C \cdot \text{CTR}_t.$$
In this case, $\delta_t^*= -\text{CVR}_t$. In the case of inactive index, $CTR_t CVR_t+C u_0 CTR_t=(\gamma + u_0)bid_t$, and the equation above is true when $bid_t \leq wp_t$ and the auction is lost.
Active indices are those for which the maximum is not opened at 0.
$$-\delta_t^* CTR_t - \gamma wp_t -  u_0 wp_t+ u_0 C \cdot \text{CTR}_t \geq 0.$$

$$ \text{CTR}_t CVR_t - \frac{\alpha CTR_t}{ \sqrt{\sum_{l \in \mathcal{T}} CTR^2_l}} CTR_t - \gamma wp_t  -  u_0 wp_t+ u_0 C \cdot \text{CTR}_t \geq 0.$$

$$ \text{CTR}_t CVR_t + u_0 C \cdot \text{CTR}_t - \alpha \frac{ CTR_t^2}{ \sqrt{\sum_{l \in \mathcal{T}} CTR^2_l}} \geq  \gamma wp_t +  u_0 wp_t .$$

In this case $\text{CTR}_t CVR_t + u_0 C \cdot \text{CTR}_t - \alpha \frac{ CTR_t^2}{ \sqrt{\sum_{l \in \mathcal{T}} CTR^2_l}}  = (\gamma + u_0) bid_t$, and the equation above is true when $bid_t \geq wp_t$ and the auction is won.

Overall, the next lemma can be formulated.

\begin{lemma}[The bid for the CVR uncertainty in MSE terms]
\label{app:lemma_mse_cvr}
    For the robust optimization problem \ref{opt:robust_optimization} the closed formula for the bid can be obtained \ref{eq::robust_bid_ctr} with active indexes defined as \ref{inactive_index_formula}
\end{lemma}

\section{Robust problem reformulation and solution for joint CTR-CVR uncertainties}
\label{sec::appendix_joint}
We consider the robust optimization problem
\[
\begin{aligned}
\max_{x\in\R^T_+}\;\min_{\substack{\sum_{t=1}^T(b_t-\CVR_t)^2\le2\varepsilon^b\\
                                   \sum_{t=1}^T(a_t-\CTR_t)^2\le2\varepsilon^a}}
\;\sum_{t=1}^T x_t\,a_t\,b_t
\quad\text{s.t.}\quad
\sum_{t=1}^T x_t\,wp_t\le B,\quad
\frac{\sum_{t=1}^T x_t\,wp_t}{\sum_{t=1}^T x_t\,a_t}\le C.
\end{aligned}
\]

\subsection{Quadratic‐form reformulation}

\begin{itemize}
  \item {\bf Decision vector:} \(x=(x_1,\dots,x_T)^\top\in\R^T_+\).
  \item {\bf Nominals:} \(a^0_t=\CTR_t,\;b^0_t=\CVR_t\).
  \item {\bf Perturbations:} 
    \(\delta a_t=a_t-a^0_t,\;\delta b_t=b_t-b^0_t\), 
    subject to 
    \(\|\delta a\|_2^2\le r_a^2=2\varepsilon^a,\;\|\delta b\|_2^2\le r_b^2=2\varepsilon^b.\)
  \item {\bf Inner objective:}
    \[
      \sum_{t=1}^T x_t\,a_t\,b_t
      =\sum_t x_t\,(a^0_t+\delta a_t)(b^0_t+\delta b_t)
      =z^\top Q\,z + 2\,q^\top z + c,
    \]
    where
    \[
      z=\begin{pmatrix}\delta a\\\delta b\end{pmatrix},\quad
      D=\diag(x_1,\dots,x_T),\quad
      Q=\begin{pmatrix}0 & \tfrac12D\\[3pt]\tfrac12D & 0\end{pmatrix},
    \]
    \[
      q=\tfrac12\begin{pmatrix}D\,b^0\\[3pt]D\,a^0\end{pmatrix},\quad
      c=\sum_{t=1}^T x_t\,a^0_t\,b^0_t.
    \]
\end{itemize}

\subsection{Lower‐bound trick}

We introduce a slack scalar \(s\) and enforce
\[
\min_{\|\delta a\|\le r_a,\|\delta b\|\le r_b}
\;(z^\top Qz +2q^\top z + c)
\;\ge\;s
\quad\Longleftrightarrow\quad
z^\top Qz + 2q^\top z + (c-s)\;\ge0
\quad\forall\,z.
\]



\subsection{ Enforcing via the S‑lemma}

Define the two “ball” inequalities
\[
g_1(z)\equiv r_a^2-\|\delta a\|^2 = r_a^2 - \delta a^T I\delta a\;\ge0,\quad
g_2(z)\equiv r_b^2-\|\delta b\|^2= r_b^2 - \delta b^T I\delta b \;\ge0.
\]
By the inhomogeneous S‑lemma (Theorem B.2.1(ii), Appendix B.2 in Ben‑Tal \& Nemirovski):
\[
g_1(z)\ge0,\;g_2(z)\ge0
\;\Longrightarrow\;
z^\top Qz +2q^\top z + (c-s)\ge0
\quad\forall z
\]
holds if and only if there exist \(\lambda_a,\lambda_b\ge0\) such that
\[
\underbrace{
\begin{pmatrix}
Q + \diag(\lambda_aI,\lambda_bI) & q\\[4pt]
q^\top & (c-s-\lambda_a r_a^2-\lambda_b r_b^2)
\end{pmatrix}
}_{\displaystyle M(\lambda_a,\lambda_b)}
\succeq0.
\]

\subsection{ Schur‑complement decomposition}

By the Schur‐complement lemma (Prop A.2.3 in Ben‑Tal \& Nemirovski’s \emph{Lectures…}),  
\(M\succeq0\) is equivalent to the two convex conditions:

\[
\underbrace{
\begin{pmatrix}
\lambda_a I_T & \tfrac12D\\[3pt]
\tfrac12D    & \lambda_b I_T
\end{pmatrix}
}_{\succeq0}
\quad\text{and}\quad
\underbrace{
(c - s -\lambda_a r_a^2 - \lambda_b r_b^2 ) - q^T (Q + \diag(\lambda_a I, \lambda_b I))^{-1} q
}_{\geq0}.
\]

By combining the fact that $(Q + \diag(\lambda_a I, \lambda_b I))^{-1} \succeq 0$, since inversion preserves positive semi-definiteness, and applying the definition, we obtain:

\[
\begin{pmatrix}
\lambda_a I_T & \tfrac12D\\[3pt]
\tfrac12D    & \lambda_b I_T
\end{pmatrix} \succeq0 
\quad\text{and}\quad
-\lambda_a r_a^2 - \lambda_b r_b^2 + c - s
\geq0.
\]
Here the first is a \(2T\times2T\) linear matrix inequality, and the second is an affine bound.

\subsection{Robust Cost‐per‐Click constraint}

Original
\(\sum x_t\,wp_t / \sum x_t\,a_t\le C\).
Under \(\|\delta a\|\le r_a\), worst‐case denominator is
\[
\min_{\|\delta a\|\le r_a}\sum_t x_t\,(a^0_t+\delta a_t)
= \sum_t x_t\,a^0_t - r_a\,\|x\|_2.
\]
Hence enforce the single second order cone constraint
\[
\sum_{t=1}^T x_t\,wp_t
\;\le\;
C\Bigl(\sum_{t=1}^T x_t\,a^0_t - r_a\,\|x\|_2\Bigr),
\]
equivalently
\(\sum x wp - C\sum x a^0 + C\,r_a\|x\|_2\le0\).

\subsection{ Final reformulation}

Collecting variables \(x\in\R^T_+\), \(t\in\R\), \(\lambda_a,\lambda_b\ge0\), the
robust counterpart is the convex program
\[
\begin{aligned}
\max_{x,s,\lambda_a,\lambda_b}\quad &s\\[4pt]
\text{s.t.}\quad
&\sum_{t=1}^T x_t\,wp_t \;\le\; B,\\[4pt]
&\sum_{t=1}^T x_t\,wp_t - C\sum_{t=1}^T x_t\,a^0_t + C\,r_a\,\|x\|_2 \;\le\;0,\\[5pt]
&\begin{pmatrix}
\lambda_a I_T & \tfrac12\,\diag(x)\\[4pt]
\tfrac12\,\diag(x) & \lambda_b I_T
\end{pmatrix}
\succeq0,\\[5pt]
&-\lambda_a\,r_a^2 - \lambda_b\,r_b^2 + \sum_{t=1}^T x_t\,a^0_t\,b^0_t - s \;\geq\;0.
\end{aligned}
\]

Then it is possible to optimize number of variables:

\[
\begin{aligned}
\max_{x_1, \ldots,x_T,\lambda_a,\lambda_b}\quad &-\lambda_a\,r_a^2 - \lambda_b\,r_b^2 + \sum_{t=1}^T x_t\,a^0_t\,b^0_t \\[4pt]
\text{s.t.}\quad
&\sum_{t=1}^T x_t\,wp_t \;\le\; B,\\[4pt]
&\sum_{t=1}^T x_t\,wp_t - C\sum_{t=1}^T x_t\,a^0_t + C\,r_a\,\|x\|_2 \;\le\;0,\\[5pt]
&\begin{pmatrix}
\lambda_a I_T & \tfrac12\,\diag(x)\\[4pt]
\tfrac12\,\diag(x) & \lambda_b I_T
\end{pmatrix}
\succeq0,\\[5pt]
&\lambda_a \geq 0, \; \lambda_b \geq 0, \; 0 \leq x_t \leq 1
\end{aligned}
\]

Fortunately, the positive semi-definite constraint can be simplified by exploiting Shnur's complement theorem.

Applying it, the following system is obtained:
\[
\begin{cases}
    \lambda_a I_T \succeq0 \\
    \lambda_b I_T - (\frac{1}{2} \diag(x))^T (\lambda_a I_T)^{-1} (\frac{1}{2} \diag(x)) \succeq 0
\end{cases}
\]
Which is equal to:

\[
\begin{cases}
    \lambda_a I_T \succeq 0 \\
    \lambda_b I_T - \frac{1}{4 \lambda_a} \diag(x) ^2 \succeq 0
\end{cases}
\]
And due to diagonal structure of all positive semi-definite matrices:

\[
\begin{cases}
    \lambda_a \geq0,\\
    \lambda_a \lambda_b \geq \frac{1}{4} x_t^2 \quad \forall t \in 1, ..., T
\end{cases}
\]

Therefore the overall problem can be expressed as:

\[
\begin{aligned}
\max_{x_1, \ldots,x_T,\lambda_a,\lambda_b}\quad &-\lambda_a\,r_a^2 - \lambda_b\,r_b^2 + \sum_{t=1}^T x_t\,a^0_t\,b^0_t - \sum_{t=1}^T \frac{2 x_t^2 [\lambda_b (b_t^0)^2 + \lambda_a (a_t^0)^2] - 2 x_t^3 a_t^0 b_t^0}{4 \lambda_a \lambda_b - x_t^2} \\[4pt]
\text{s.t.}\quad
&\sum_{t=1}^T x_t\,wp_t \;\le\; B,\\[4pt]
&\sum_{t=1}^T x_t\,wp_t - C\sum_{t=1}^T x_t\,a^0_t + C\,r_a\,\|x\|_2 \;\le\;0,\\[5pt]
&\lambda_a \lambda_b \geq \frac{1}{4} x_t^2,\\[5pt]
&\lambda_a \geq 0, \; \lambda_b \geq 0, \; 0 \leq x_t \leq 1
\end{aligned}
\]

And then Lagrangian function is written as:

\begin{eqnarray*}
    \mathbb{L}(x_1, \ldots, x_T, \lambda_a, \lambda_b, \gamma, \mathbf{u}, -u_0, (\xi_1)_1, \ldots, (\xi_1)_T, (\xi_2)_1, \ldots, (\xi_2)_T, \zeta_a, \zeta_b) =\\ \sum_{t=1}^T x_t a_t^0 b_t^0 - \lambda_a r_a^2 - \lambda_b r_b^2 - \gamma (\sum_{t=1}^T x_t wp_t - B) + \\ (\mathbf{u}, u_0)
     \begin{pmatrix}
        \alpha\mathbf{x} \\ -\sum_{t=1}^T x_t wp_t + C \sum_{t=1}^T x_t a_t^0
    \end{pmatrix} + \sum_{t=1}^T x_t (\xi_1)_t - \sum_{t=1}^T (\xi_2)_t (x_t - 1) +\\ \zeta_a \lambda_a + \zeta_b \lambda_b - f(t, \lambda_a, \lambda_b)= \\
\end{eqnarray*}

From complementary slackness we can obtain next constraint for the optimal solution:

\[
\begin{cases}
    \frac{\partial \mathbb{L}}{\partial \lambda_a} = -r_a^2 + \zeta_a + \lambda_b - \frac{\partial f}{\partial \lambda_a} = 0 \\
    \frac{\partial \mathbb{L}}{\partial \lambda_b} = -r_b^2 + \zeta_b + \lambda_a - \frac{\partial f}{\partial \lambda_b} = 0 \\
    \zeta_a \lambda_a = 0 \\
    \zeta_b \lambda_b = 0
\end{cases}
\]

Where $\frac{\partial f}{\partial \lambda_a} = \sum_{t=1}^T \frac{(4 \lambda_a \lambda_b - x_t^2) 2 x_t^2 CTR_t^2 - 4 \lambda_b(2x_t^2 \left[ \lambda_b CVR_t^2 + \lambda_a CTR_t^2 \right] - 2x_t^3 CTR_t CVR_t)}{4 \lambda_a \lambda_ - x_t^2}$ and

$\frac{\partial f}{\partial \lambda_b} = \sum_{t=1}^T \frac{(4 \lambda_a \lambda_b - x_t^2) 2 x_t^2 CVR_t^2 - 4 \lambda_a(2x_t^2 \left[ \lambda_b CVR_t^2 + \lambda_a CTR_t^2 \right] - 2x_t^3 CTR_t CVR_t)}{4 \lambda_a \lambda_ - x_t^2}$

As optimal $\lambda_a \neq 0$ and $\lambda_b \neq 0$, the exact formulas for the $\lambda_a, \lambda_b$. Also, the fact that $x_t \geq 0$ can simplify constraint to $x_t \leq 2 \; \sqrt{\lambda_a, \lambda_b}$, thus it will be satisfied by solving the previous system and can be skipped.

Therefore, the problem is simplified or for known values of $\lambda_a$ and $\lambda_b$:

\[
\begin{aligned}
    \max_{x_1, \ldots,x_T}\quad 
    & \sum_{t=1}^T x_t\,CTR_t\,CVR_t - f(x_t, \lambda_a^*, \lambda_b^*) \\[4pt]
    \text{s.t.}\quad
    &\sum_{t=1}^T x_t\,wp_t \;\le\; B,\\[4pt]
    &\sum_{t=1}^T x_t\,wp_t - C\sum_{t=1}^T x_t\,a^0_t + C\,r_a\,\|x\|_2 \;\le\;0,\\[5pt]
    & 0 \leq x_t \leq 1
\end{aligned}
\]

Now the Lagrangian can be updated:

\begin{eqnarray*}
    \mathbb{L}(x_1, \ldots, x_T, \gamma, \mathbf{u}, -u_0, (\xi_1)_1, \ldots, (\xi_1)_T, (\xi_2)_1, \ldots, (\xi_2)_T) = \sum_{t=1}^T x_t a_t^0 b_t^0 - f(x_t, \lambda_a^*, \lambda_b^*)  - \gamma (\sum_{t=1}^T x_t wp_t - B) + \\ (\mathbf{u}, u_0)
     \begin{pmatrix}
        \alpha\mathbf{x} \\ -\sum_{t=1}^T x_t wp_t + C \sum_{t=1}^T x_t a_t^0
    \end{pmatrix} + \sum_{t=1}^T x_t (\xi_1)_t - \sum_{t=1}^T (\xi_2)_t (x_t - 1) = \\
    \sum_{t=1}^T x_t (a_t^0 b_t^0 - \gamma wp_t + \alpha u_t + u_0 (C \cdot CTR_t - wp_t) + (\xi_1)_t - (\xi_2)_t ) +
    \gamma B + \sum_{t=1}^T (\xi_2)_t
\end{eqnarray*}

Now the dual problem can be constructed (also add an optimality condition to the problem):

\begin{eqnarray*}
    \min_{\gamma, \mathbf{u}, u_0, \mathbf{\xi_1}, \mathbf{\xi_2}} \gamma B + \sum_{t=1}^T (\xi_2)_t  \\
    \text{s.t.} \: a_t^0 b_t^0 - \gamma wp_t + u_0 (-wp_t + C \cdot CTR_t) + u_t \alpha + (\xi_1)_t - (\xi_2)_t = \frac{\partial f}{\partial x_t}(x_t, \lambda_a, \lambda_b)= \\
    \frac{ (4 \lambda_a \lambda_b - x_t^2) \; (4 x_t[\lambda_a CTR_t^2 + \lambda_b CVR_t^2] - 6 x_t^2 CTR_t CVR_t) \; + \; 4 x_t (x_t^2 \lambda_a CTR_t^2 + x_t^2 \lambda_b CVR_t^2 - x_t^3 CTR_t CVR_t)}{(4 \lambda_a \lambda_b - x_t^2)^2}\\
    ||\mathbf{u}||_2 \leq u_0, \; (\xi_1)_t \geq 0, \; (\xi_2)_t \geq 0,
\end{eqnarray*}

Using similar tricks as for the robust MSE with uncertainty of $CTR_t$, the optimization is transformed to:

\begin{eqnarray*}
    \min_{\gamma, \mathbf{u}, u_0, \mathbf{\xi_1}, \mathbf{\xi_2}} \gamma B + \sum_{t=1}^T \max(0, a_t^0 b_t^0 - \gamma wp_t + u_0 (-wp_t + C \cdot CTR_t) + u_t \alpha - \frac{\partial f}{\partial x_t}\big\vert_{x_t = 1} )  \\
    \text{s.t.} \:  ||\mathbf{u}||_2 \leq u_0 \\
    \frac{\partial f}{\partial x_t}\big\vert_{x_t = 1} =\frac{ (4 \lambda_a \lambda_b - 1) \; (4[\lambda_a CTR_t^2 + \lambda_b CVR_t^2] - 6 CTR_t CVR_t) \; + \; 4(\lambda_a CTR_t^2 + \lambda_b CVR_t^2 - CTR_t CVR_t)}{(4 \lambda_a \lambda_b - 1)^2}
\end{eqnarray*}

And the resulting formula for bid will be equal to 

\begin{multline*}   
bid_t = \frac{CTR_t CVR_t}{\gamma +u_0} +\frac {u_0 C \cdot CTR_t} {\gamma+u_0}+
\begin{cases}
     - \frac {\alpha}{\gamma+u_0} \left( \frac{u_0}{\sqrt{\mid\mathcal{T}\mid}} + \frac{\partial f}{\partial x_t}\big\vert_{x_t = 1} \right) \quad, t \in \mathcal{T};\\
     0,
    \quad t \notin \mathcal{T}.
\end{cases}
\end{multline*}

where $\frac{\partial f}{\partial x_t}\big\vert_{x_t = 1} =\frac{ (4 \lambda_a \lambda_b - 1) \; (4[\lambda_a CTR_t^2 + \lambda_b CVR_t^2] - 6 CTR_t CVR_t) \; + \; 4(\lambda_a CTR_t^2 + \lambda_b CVR_t^2 - CTR_t CVR_t)}{(4 \lambda_a \lambda_b - 1)^2}$

Key reference for S-lemma, Schnur complement theorem and   - \cite{book_robustopt}

Overall, the next lemma about the solution can be formulated:

\begin{lemma}[The bid for the CTR and CVR uncertainty in MSE terms]
\label{app:lemma_mse_ctr_cvr}
    For the robust optimization problem \ref{opt:joint_mse_problem} the closed formula for the bid can be obtained \ref{bid_Joint_MSE_main} with active indexes defined as \ref{inactive_index_formula_joint}.
\end{lemma}

\label{sec::appendix-nonrobust}
This section presents the optimization problem used as the baseline solution in the experiments. It was originally introduced in \cite{pid2019}. The problem leverages the same constraints (Budget and Cost-Per-Click):
\begin{align*}
&\max _{x_{t}} \sum _{t=1}^T x_{t} \cdot CTR_{t} \cdot CVR_{t} \\
&\text { s.t. } \quad \sum _{t=1}^T x_{t} \cdot w p_{t} \leq B \\
& \hspace{1cm} \frac{\sum_{t=1}^T x_{t} \cdot w p_{t}}{\sum_{t=1}^T x_{t} \cdot C T R_{t}} \leq C \\
&\text { where } \quad 0 \leq x_{t} \leq 1, \quad \forall t \in [1, T].
\label{eq::non-robust}
\end{align*}

Then the dual problem is formulated:

\begin{equation*}
\begin{array}{ll}
\min _{p, q, r} & B \cdot p + \sum_{t=1}^{N} r_t \\
\text{s.t.} & w p_t \cdot p + \left(w p_t - C T R_t \cdot C\right) q + r_t \geq v_t, \quad \forall i \\
\text{where} & p \geq 0, \\
& q \geq 0, \\
& r_t \geq 0, \quad \forall t \\
& v_t = C T R_t \cdot C V R_t, \quad \forall t.
\end{array}
\end{equation*}

Finally, the total formula for bid will be as follows:  $$bid_t = \frac{1}{p+q} CVR_t CTR_t + \frac{q}{p + q} C \cdot CTR_t.$$

Where $p, q$ corresponds to the Budget and CPC constraint. 

Or in terms of the solution from the paper:

$$bid_t = \frac{1}{\gamma + u_0} CVR_t CTR_t + \frac{\gamma}{\gamma + u_0} C \cdot CTR_t, $$
where $\gamma, u_0)$ corresponds to Budget and CPC constraint. While $\varepsilon_a \xrightarrow{} 0$ then $\alpha \xrightarrow{} 0 $ and therefore the additional summand from equations \ref{eq::robust_bid_ctr} and \ref{bid_Joint_MSE_main} converges to $0$ and the total formulas transforms to~\cite{pid2019}.
Thus, the received formulas match the classic solution.

\section{Experiment details}
\label{sec::appendix-experiments}

The optimization process consists of two main stages. First, initial values for dual parameters are manually selected. 
On the second step, with some historical data accumulated, these parameters are refined dynamically using the ECOS solver \cite{domahidi2013ecos}, which optimizes robust formulation (see equation ~\eqref{robust_formulation_exp}) at each auction step. 

To simulate uncertainty in CTR predictions, multiple uncertainty levels $\varepsilon$ are established, introducing perturbations to the actual CTR distribution such that $L(\bctr^{\text{noised}}, \bctr^i) \leq \varepsilon$. 
Experimental results for each $\varepsilon$ are averaged across 10 random seeds.

In the MSE and MAE normalizations, the variation of $\varepsilon$ is naturally carried out from 0 to permissible upper boundary. 
In the CE normalization, the correlation of true $CTR$ values (even with zero noise) is essentially non-zero (this value is easy to obtain analytically, knowing the limits of acceptable $CTR$). By varying the $CTR$ within the same limits, a variation within the range of permissible $\varepsilon$ is obtained. 

Table ~\ref{table:parameters} shows budget ($B$) and cost-per-click ($C$) constraint values for each experimental setting. Values for the BAT setup are derived directly from the real advertising campaigns data distribution. All three experimental setups use the same cold-start bidding strategy: relatively low initial bids to ensure not to spend all its' budget in the beginning.

\begin{table}[h]
  \caption{B and C constraints for each experiment set.}  
  \label{table:parameters}
  \centering
  \begin{tabular}{lccc}
    \toprule
    \textbf{Parameter} & \textbf{Synthetic} & \textbf{iPinYou} & \textbf{BAT} \\
    \midrule
    B (budget)         & 1                  & 500             & $[4, 6700]$  \\
    C (constraint)     & 1                  & 500             & 300          \\
    \bottomrule
  \end{tabular}
\end{table}

The ratio between $B$ and $C$ is associated with usually low CTR values in production and as a consequence synthetic simulation. From the formula \eqref{opt:robust_optimization} for calculating the average cost-per-click, it follows that the bid in average should not exceed $CTR_t\cdot C$. Budget range for the BAT setup is derived directly from the real advertising campaigns data distribution. 

In the synthetic and iPinYou experiments, the auction results and click events rely on the true $CTR_t^i$ and $CVR_t^i$ values, while in the BAT experiment, they are derived from historical bid-dependent patterns.

In BAT protocol, each advertising campaign operates independently, competing with others based on aggregated historical data. For each bid, the advertiser is charged either the full bid amount or its remaining budget (if less than the bid). Due to data aggregation specifics, clicks and conversions are allocated based on bid amounts rather than determining explicit auction winners. The corresponding CTR and CVR values are selected from historical distributions. Campaign performance is tracked using a click-based threshold to define winning status, which is necessary for MSE calculations.

To demonstrate the statistical significance of the uncertainty in $CTR$ and $CVR$, we present the standard deviation metrics in Table~\ref{tab::std_mse_ctrcvr}.

\begin{table}[!h]
\caption{Standard deviation of the experiments with MSE loss under dual uncertainty (CTR and CVR).}
\begin{tabular}{lcccccc}
\toprule
Metric & \multicolumn{2}{c}{Synthetic} & \multicolumn{2}{c}{IPinYou} & \multicolumn{2}{c}{BAT} \\
\cmidrule(lr){2-3} \cmidrule(lr){4-5} \cmidrule(lr){6-7}
 & Robust & Non-robust & Robust & Non-robust & Robust & Non-robust \\
\midrule
$TVC$ std & 0.02 & 0.01 & 0.004 & 0.006 & 0.01 & 0.01 \\
$CPC\_avg$ std & 0.32 & 0.03 & 25.11 & 30.69 & 0.17 & 4.87 \\
\bottomrule
\end{tabular}
\label{tab::std_mse_ctrcvr}
\end{table}

\subsection*{Compute Resources}

All experiments in this paper were conducted on the M2 Pro CPU with 32 GB of Memory. Each experiment for each algorithm does not last more than 2 hours of computation. Therefore, all results from this paper can be reproduced on the average PC. 

\label{sec:appendix_old_metrics}

\begin{figure*}[t]
    \centering
    
     \begin{tabular}{p{0.31\textwidth}p{0.31\textwidth}p{0.31\textwidth}}
        \centering MSE  & \centering CE 
    \end{tabular}

     \begin{tabular}{cc}
    \includegraphics[width=0.31\textwidth]{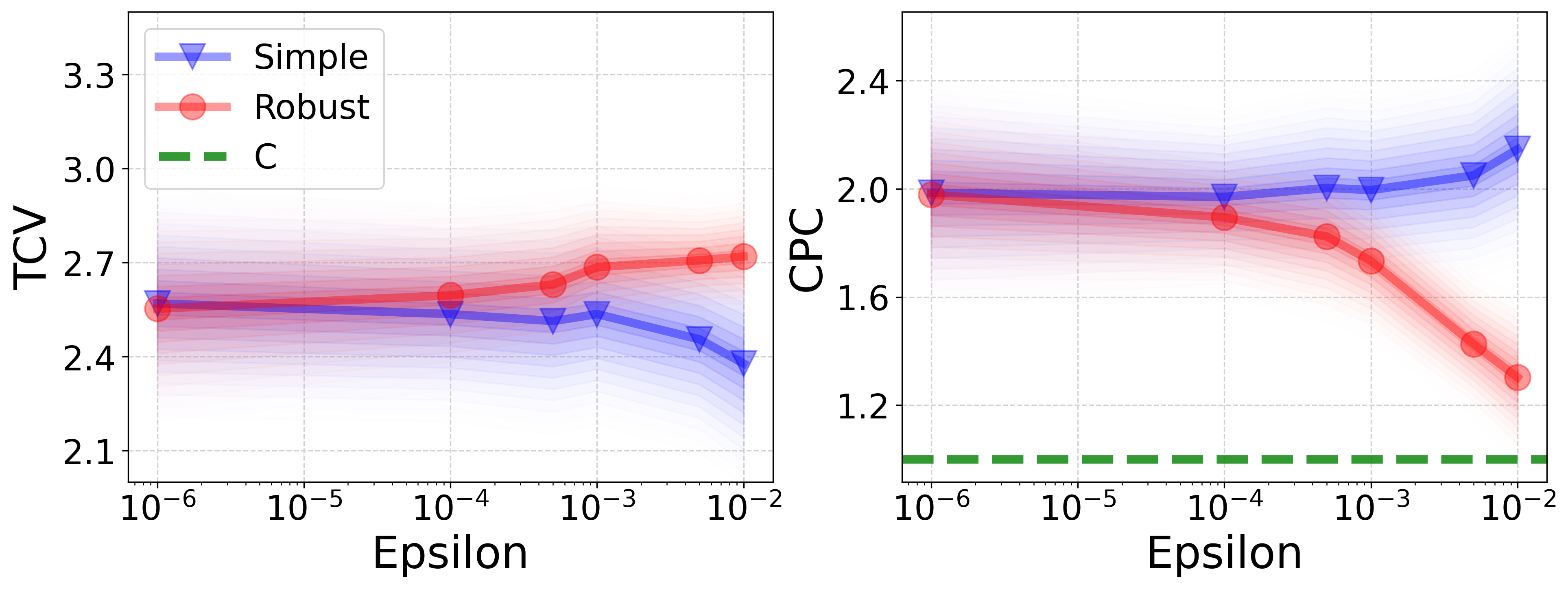} &
    \includegraphics[width=0.31\textwidth]{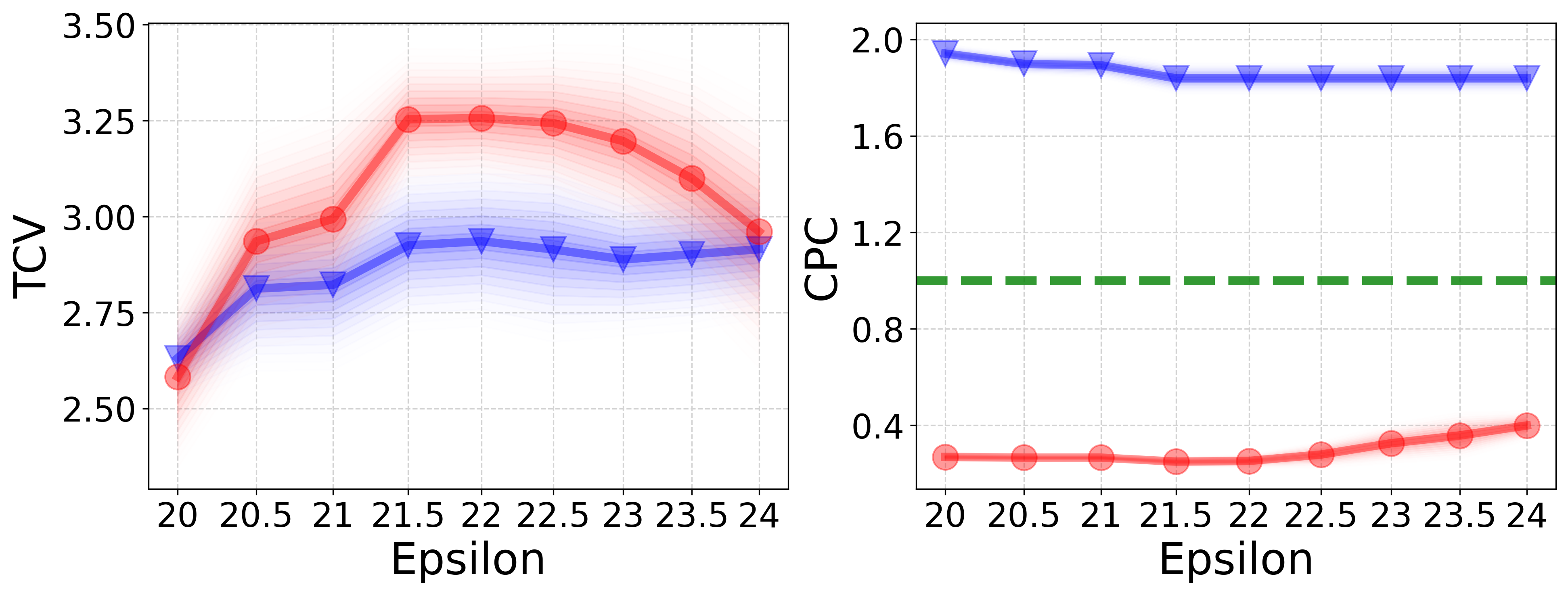} \\
    \multicolumn{2}{c}{Synthetic experiment} \\[5pt]
    
    \includegraphics[width=0.31\textwidth]{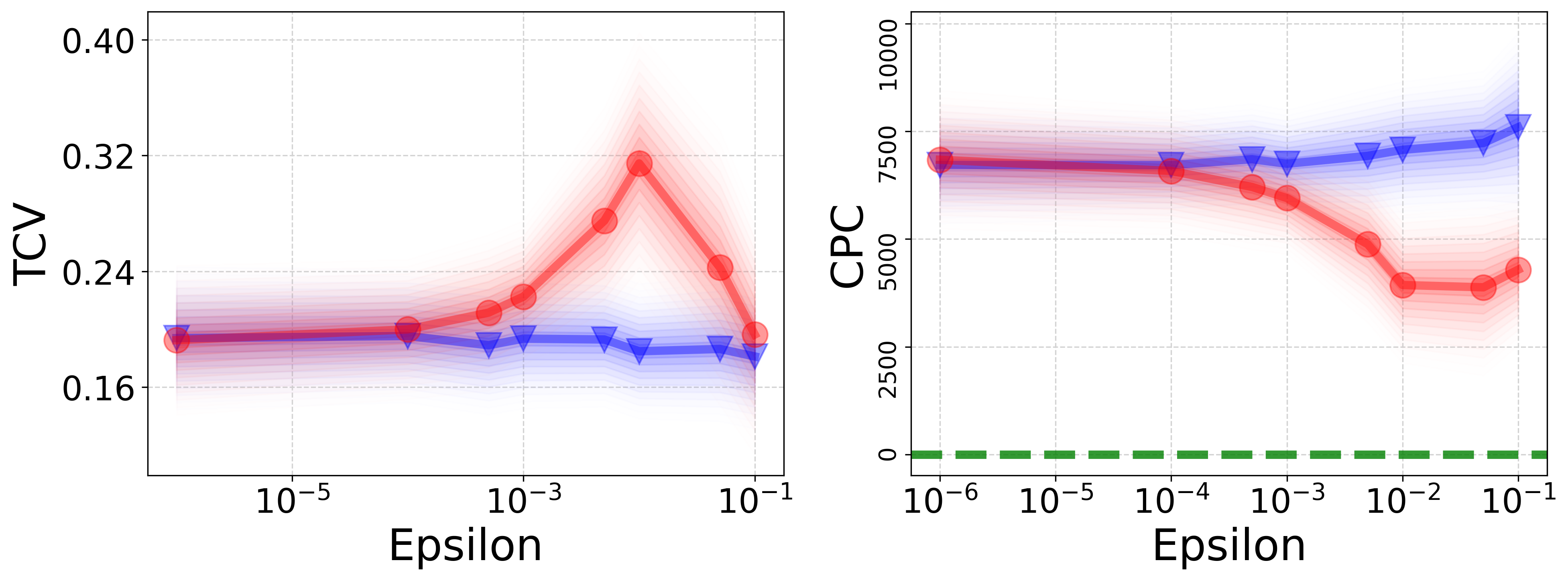} &
    \includegraphics[width=0.31\textwidth]{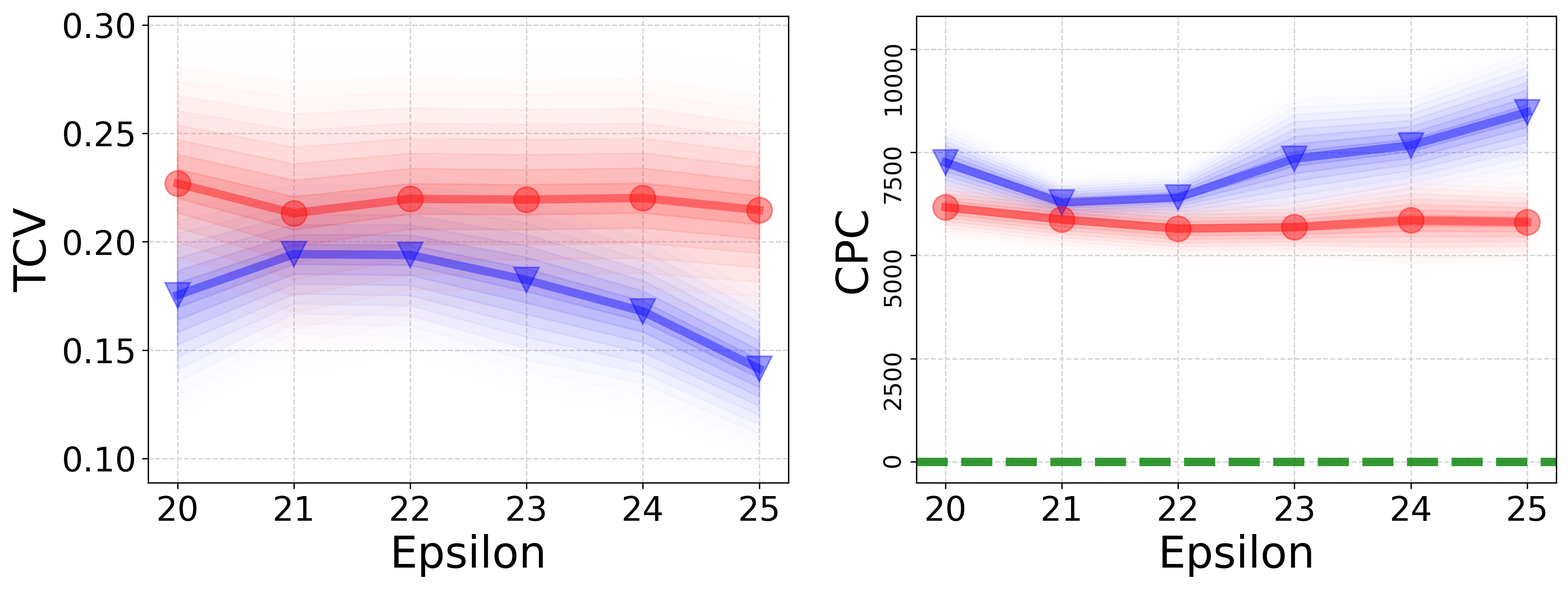} \\
    \multicolumn{2}{c}{IPinYou experiment} \\[5pt]
    
    \includegraphics[width=0.31\textwidth]{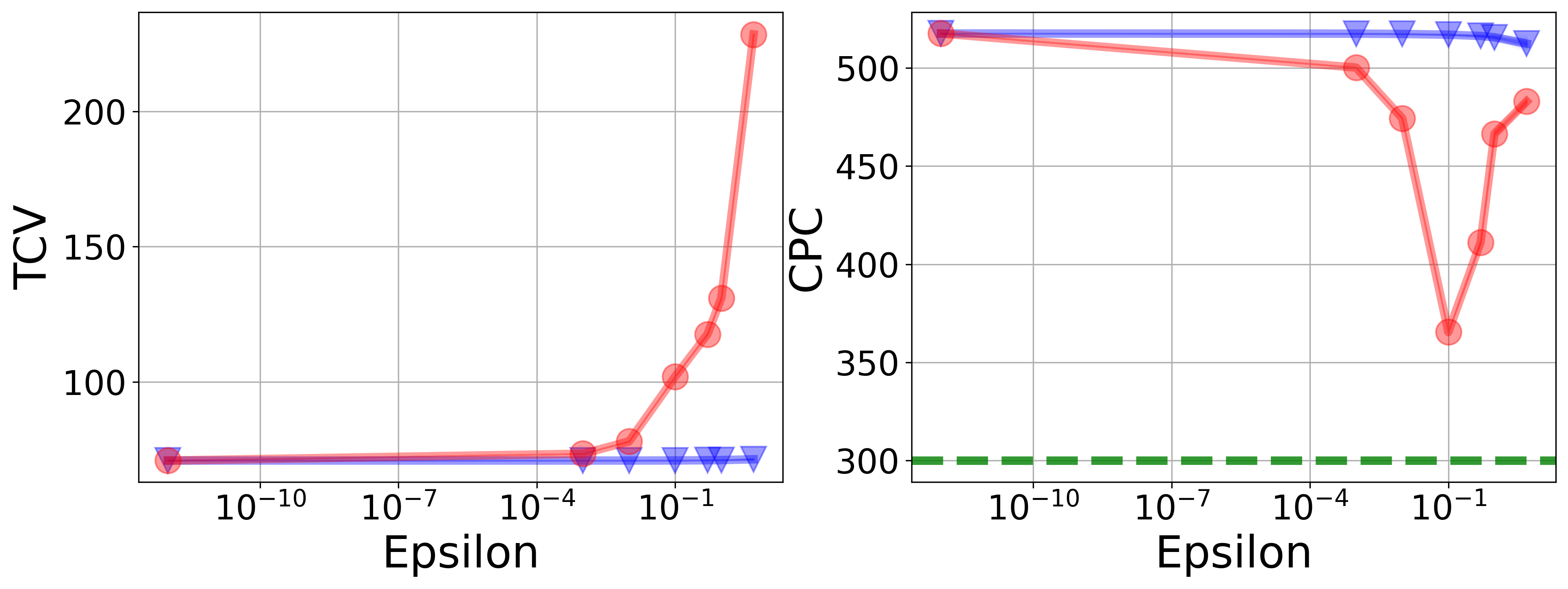} &
    \includegraphics[width=0.31\textwidth]{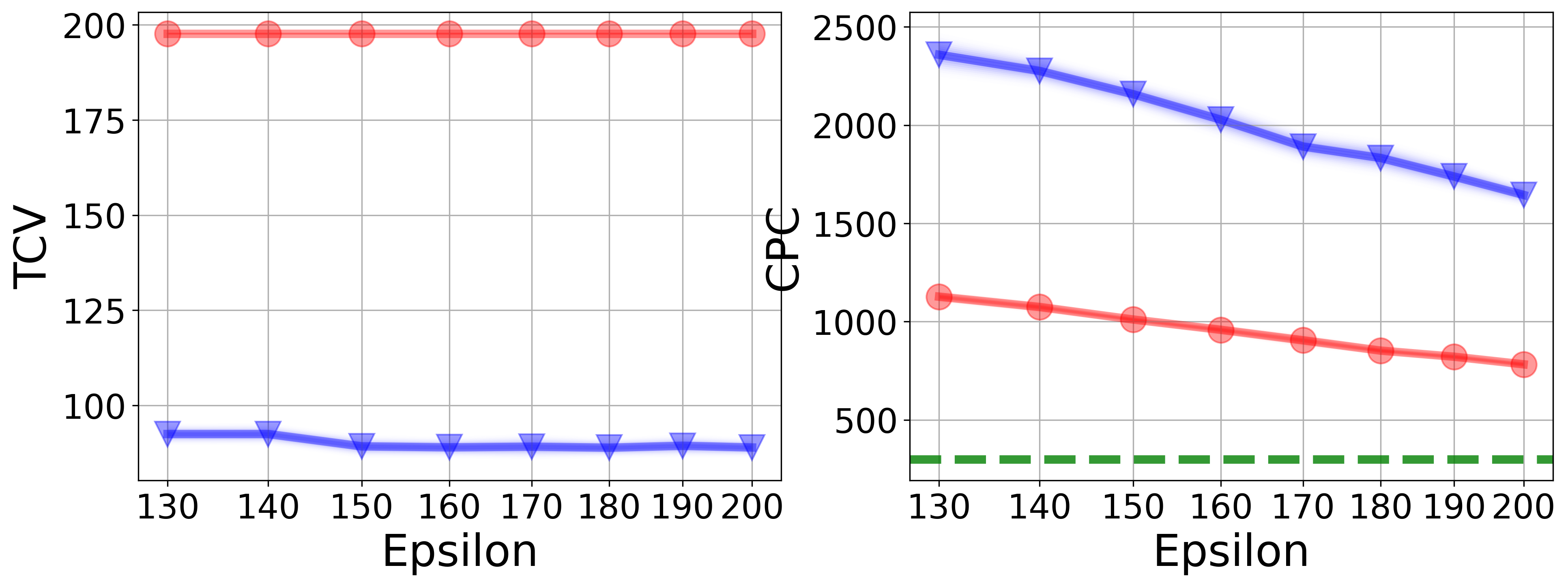} \\
    \multicolumn{2}{c}{BAT experiment}
    \end{tabular}
    
    \caption{Experimental results. The blue line indicates the value of the metrics on the non-robust solution of the problem, the red line - on the robust one. The green dotted line represents the CPC limit. Shades of red specify the root mean square error in the gradient shading form.}
    \label{fig:all_results}
\end{figure*}

The dependencies of the metrics on the $CTR$ uncertainty level,assuming CVR is known exactly, for MSE, MAE and CE 
error estimates are presented in the Figure ~\ref{fig:all_results}. Note that with zero epsilon, the results of the robust and non-robust methods naturally converge to the same values in case of MSE loss, but in other cases they do not since the formulas of the robust and non-robust algorithms do not converge for ideal CTR prediction.

In all experiments, the total conversions TCV when using the robust algorithm are guaranteed to exceed conversions using the non-robust one.  The advantage of the robust method with respect to this metric can be stable (in cases of BAT MAE, iPinYou and BAT CE experiments) or increase with increasing error in the CTR estimate (in cases of Synthetic, IpinYou and BAT MSE, Synthetic and IpinYou MAE, Synthetic CE), and further convergence of the metric values can be observed when approaching the maximum epsilon at which the optimization system is solvable (Synthetic and IPinYou MSE, Synthetic, IPinYou and BAT MAE, Synthetic CE).

\end{document}